\documentclass[english,prb,aps,amsmath,amssymb,reprint,superscriptaddress,showkeys]{revtex4-2}
\usepackage{mathptmx}
\usepackage[T1]{fontenc}
\usepackage[utf8]{inputenc}
\usepackage{color}
\usepackage{babel}
\usepackage{textcomp}
\usepackage{amsmath}
\usepackage{amssymb}
\usepackage{graphicx}
\usepackage{wasysym}
\usepackage[unicode=true,
 bookmarks=true,bookmarksnumbered=true,bookmarksopen=true,bookmarksopenlevel=2,
 breaklinks=true,pdfborder={0 0 1},backref=false,colorlinks=true]
 {hyperref}
\hypersetup{pdftitle={Theory of reactions between hydrogen and group-III acceptors in silicon},
 pdfauthor={José Coutinho},
 allcolors=blue}

\makeatletter

\providecommand{\tabularnewline}{\\}

\usepackage{orcidlink}

\makeatother

\begin{document}
\title{Theory of reactions between hydrogen and group-III acceptors in silicon}
\author{José Coutinho\orcidlink{0000-0003-0280-366X}}
\affiliation{I3N, Department of Physics, University of Aveiro, Campus Santiago,
3810-193 Aveiro, Portugal}
\email{jose.coutinho@ua.pt}

\author{Diana Gomes\orcidlink{0000-0001-8263-8523}}
\affiliation{I3N, Department of Physics, University of Aveiro, Campus Santiago,
3810-193 Aveiro, Portugal}
\author{Vitor J. B. Torres\orcidlink{0000-0003-3795-6818}}
\affiliation{I3N, Department of Physics, University of Aveiro, Campus Santiago,
3810-193 Aveiro, Portugal}
\author{\linebreak{}
Tarek O. Abdul Fattah\orcidlink{0000-0001-8734-6706}}
\affiliation{Photon Science Institute and Department of Electrical and Electronic
Engineering, The University of Manchester, Manchester M13 9PL, United
Kingdom}
\author{Vladimir P. Markevich\orcidlink{0000-0002-2503-6144}}
\affiliation{Photon Science Institute and Department of Electrical and Electronic
Engineering, The University of Manchester, Manchester M13 9PL, United
Kingdom}
\author{Anthony R. Peaker\orcidlink{0000-0001-7667-4624}}
\affiliation{Photon Science Institute and Department of Electrical and Electronic
Engineering, The University of Manchester, Manchester M13 9PL, United
Kingdom}
\begin{abstract}
The thermodynamics of several reactions involving atomic and molecular
hydrogen with group-III acceptors is investigated. The results provide
a first-principles-level account of thermally- and carrier-activated
processes involving these species. Acceptor-hydrogen pairing is revisited
as well. We present a refined physicochemical picture of long-range
migration, compensation effects, and short-range reactions, leading
to fully passivated $\equiv\textrm{Si-H}\cdots X\equiv$ structures,
where $X$ is a group-III acceptor element. The formation and dissociation
of acceptor-H and acceptor-H$_{2}$ complexes is considered in the
context of Light and elevated Temperature Induced Degradation (LeTID)
of silicon-based solar cells. Besides explaining observed trends and
answering several fundamental questions regarding the properties of
acceptor-hydrogen pairing, we find that the BH$_{2}$ complex is a
by-product along the reaction of H$_{2}$ molecules with boron toward
the formation of BH pairs (along with subtraction of free holes).
The calculated changes in Helmholtz free energies upon the considered
defect reactions, as well as activation barriers for BH$_{2}$ formation/dissociation
(close to $\sim1$~eV) are compatible with the experimentally determined
activation energies of degradation/recovery rates of Si:B-based cells
during LeTID. Dihydrogenated acceptors heavier than boron are anticipated
to be effective-mass-like shallow donors, and therefore, unlikely
to show similar non-radiative recombination activity.

\noindent \emph{Pre-print published in Physical Review B 108, 014111
(2023)}

\noindent DOI:\href{https://doi.org/10.1103/PhysRevB.108.014111}{10.1103/PhysRevB.108.014111}

\end{abstract}
\maketitle

\section{Introduction\label{sec:intro}}

Among all impurities studied in crystalline silicon, hydrogen has
attracted much attention, if not the most \citep{Stutzmann1989,Pankove1991,Pearton1992,Nickel1999,Estreicher2014b}.
Isolated atomic hydrogen in Si shows amphoteric character, high mobility
as well as the ability to form strong bonds with other elements. This
combination of properties has made H especially useful in many bulk
\citep{Pankove1983,Pearton1987,Mullins2017} and surface \citep{Seager1985,Sopori1996,Aberle2000}
engineering processes, with profound impact on device performance.

The generally accepted picture of isolated H in Si is that of a species
that can trap a hole to become a proton located at the center of a
Si-Si bond. However, it can also trap an electron to form a H$^{-}$
ion, which according to theorists, finds its most stable state at
(or close to) the tetrahedral interstitial site \citep{Deak1988,VandeWalle1989,Sasaki1989,Jones1991,Herring2001,Estreicher2012}.
The bistability of H is at the origin of the negative-$U$ of its
neutral state, which disproportionates into ionic species, $\textrm{H}^{0}\rightarrow x\textrm{H}^{+}+(1-x)\textrm{H}^{-}$,
with the fractional concentration $x$ depending on the location of
the Fermi level with respect to the $(-/+)$ transition level \citep{VandeWalle1989},
estimated experimentally in the range 0.3-0.4~eV below the conduction
band bottom \citep{Johnson1994,Johnson1995,Nielsen2002}.

Hydrogen-related defect engineering has conspicuous applications in
the solar industry. Generally, Si solar cells feature a hydrogen-rich
silicon nitride (SiN$_{x}$:H) or oxide layers deposited over the
surface, providing chemical passivation, external anti-reflection
to sunlight, as well as a highly-reflective interface to maximize
the length of internal light paths. After SiN$_{x}$:H deposition,
a step known as \emph{fast-firing} (essentially a 750-850~°C annealing
for a few seconds), induces the injection of substantial quantities
of hydrogen from the passivating layer into the silicon \citep{Jiang2003,Sheoran2008},
improving the quality of the interface, and further suppressing bulk
and surface recombination events.

The presence of H in Si is also crucial for preventing the boron-oxygen-related
light-induced degradation (BO-LID) of Si solar cells \citep{Herguth2008,Wilking2013,Helmich2021}.
BO-LID is related to a decrease of minority carrier lifetime of cells
based on B-doped O-rich silicon upon exposure to sunlight in a time
scale of hours \citep{Schmidt2004,Niewelt2017}. It has been found
that Si:(B+O)-based devices that were fired and subsequently annealed
with illumination or with minority carrier injection (a process originally
referred to as regeneration \citep{Herguth2008}), do not suffer from
BO-LID. An examination of the BO-LID and regeneration kinetics in
boron-doped p-type Czochralski-grown wafers, shows a clear dependence
on the hydrogen content with a linear increase of the regeneration
rate constant with increasing the bulk hydrogen concentration \citep{Helmich2021}.
Regeneration of solar Si was recently proposed to result from H-passivation
of the boron-dioxygen complex (BO$_{2}$) due to relocation of H from
dissociated boron-hydrogen (BH) pairs \citep{Fattah2022}. Although
it is generally thought that light can act as a catalyst for BH dissociation,
it is not clear how that works.

Hydrogen in silicon has been found to be responsible for detrimental
effects as well. Perhaps the most paradigmatic is the passivation
of dopants upon H-acceptor and H-donor pair formation \citep{Sah1983,Pankove1983,Johnson1986,Denteneer1989,Estreicher1991}.
Hydrogen-related reactions seem to be intimately related to yet another
lifetime degradation effect. This is known as Light- and elevated-Temperature-Induced
Degradation (LeTID) of solar Si \citep{Ramspeck2012}. Accordingly,
photovoltaic modules show a substantial decrease of conversion efficiency
(up to 16\% relative) due to carrier lifetime shortening, and for
that reason, it is currently an active topic of research. The LeTID
designation was originally coined as to reflect the fact that light
exposure and heat are needed for its observation, but the presence
of hydrogen was subsequently deemed necessary as well \citep{Bredemeier2019}.
Typically, the effect is manifested as a drop of the carrier lifetime
upon field-prolonged operation (time-scale of years) of solar cells
followed by a recovery. However, accelerated lab conditions, either
subjecting the cells to $\sim100$~°C anneals with $\sim1$~kW/m$^{2}$
(one sun) of illumination which is mostly above band gap, or passing
a current to inject minority carriers equivalent to the population
produced during normal operation, result in a maximum drop of the
lifetime within hours. The latter procedure is nowadays almost universally
used for LeTID testing, and for that reason, sometimes the effect
is referred to as \emph{carrier induced degradation}.

Dark anneals were also found to reproduce the LeTID effect, although
at a considerably slower rate than with illumination, allowing for
discrimination of LeTID from BO-LID. At $T=140$~°C, maximum degradation
was attained after more than 100~hours \citep{Vargas2019}. From
this work, activation energies of 1.08~eV for degradation and 1.11~eV
for recovery were measured during prolonged dark-annealing treatments.
Importantly, and despite having been observed in several materials
(multi-crystalline, Czochralski and floating-zone grown, p- and n-type),
the firing step was found to be an indispensable ingredient for LeTID.
This led to the conclusion that hydrogen is a constituent or somehow
involved in the formation of the defect responsible for the non-radiative
recombination of photogenerated carriers (see Ref.~\citep{Chen2020a}
for a review of LeTID of Si solar cells).

During the high-temperature firing step, the hydrogen introduced into
the Si is only stable in the form of ions, which become distributed
across the whole thickness of the wafers \citep{Sheoran2008}. However,
after cooling to room temperature, most of the hydrogen in the Si
bulk forms H$_{2}$ molecules, either isolated or trapped next to
impurities or defects \citep{Pritchard1998,Pritchard1999}. This is
the usual state of hydrogen in \emph{as-fired} (pre-degraded) cells.

Isolated H$_{2}$ molecules in crystalline Si start to migrate just
above 300~K. However, in O-rich material they are weakly bound to
interstitial O atoms (with a binding energy of 0.28~eV), so that
one needs to raise the temperature up to around 70~°C to initiate
molecular motion \citep{Markevich1998}. Upon annealing boron-doped
Si above 160~°C, the molecules are known to interact effectively
with boron atoms and form BH pairs. This reaction has been monitored
either by local vibrational mode (LVM) infra-red absorption spectroscopy
\citep{Pritchard1999,Weiser2020} or by changes in resistivity \citep{Voronkov2017,Walter2019}.

Recently, several groups have been investigating possible correlations
between the LeTID degradation and the evolution of hydrogen, most
notably in boron-doped Si, by monitoring the formation/dissociation
of BH pairs \citep{Fung2018,Winter2021,Walter2022,Hammann2023,Kwapil2023}.

Upon subjecting as-fired samples to isochronal annealings at several
temperatures, Fung \emph{et~al.} \citep{Fung2018} found that the
maximum LeTID defect density correlated with the BO-LID regeneration
rate (recently suggested to reflect the formation of a BO$_{2}$-H
complex \citep{Fattah2022}), as well as with formation of BH pairs.
Similar observations were reported by Hammann \emph{et~al.} \citep{Hammann2021,Hammann2023}
who found a correlation between the LeTID defect formation kinetics
and changes in resistivity, assumed to be due to BH formation, with
both quantities reaching their maximum values at around the same time.

Winter \emph{et~al.} \citep{Winter2021} used the three-state-model
of Voronkov and Falster \citep{Voronkov2017},

\begin{equation}
\textrm{H}_{2\textrm{A}}+2\textrm{B}^{-}+2\textrm{h}^{+}\rightleftarrows2\textrm{HB}\rightleftarrows\textrm{H}_{2\textrm{C}}+2\textrm{B}^{-}+2\textrm{h}^{+},\label{eq:vfmodel}
\end{equation}
to describe the dynamics of free-hole concentration (decrease followed
by recovery) in floating-zone Si subject to surface passivation/firing
and dark annealings. In the above model, H$_{2\textrm{A}}$ and H$_{2\textrm{C}}$
stand for two different types of H-dimers. The first refers to molecular
hydrogen at tetrahedral interstitial sites of the Si lattice \citep{Pritchard1998,Pritchard1999}.
The second dimer, although unidentified, must be a more stable state
than that of the molecules \citep{Voronkov2017}. It should be noted
that according to the model presented in Ref.~\citep{Voronkov2017},
the first step (left-side) of reaction~\ref{eq:vfmodel} initiates
with the dissociation of H$_{\textrm{2A}}$ upon its interaction with
a hole. The resulting positively charged H atoms then interact effectively
with negatively charged B atoms to form BH pairs. From the experimental
data it was possible to extract activation energies for formation
and dissociation of the BH pairs as 1.29\textpm 0.07 eV and 1.22\textpm 0.18
eV \citep{Winter2021}, respectively. These figures were attributed
to the rate-limiting processes of first (left-side) and second (right-side)
steps of reaction~\ref{eq:vfmodel} toward the right direction.

More recently, from prolonged dark annealings of SiN$_{x}$:H-coated/fired
structures, Walter and co-workers \citep{Walter2022} found respectively
transient and equilibrium activation energies of $0.84\pm0.17$~eV
and $1.35\pm0.15$~eV for the first step of reaction \ref{eq:vfmodel}.
The first quantity refers to the same property that was reported in
Ref.~\citep{Winter2021} as being $1.29\pm0.07$~eV. A third measurement
of this quantity was carried out by Acker \emph{et~al.} \citep{Acker2022}
and gave $1.20\pm0.02$~eV. This point needs clarification, not only
because the values derived differ by more than their error bars added
together, but also because they are close to $1.08\pm0.05$~eV that
was measured by Vargas and co-workers \citep{Vargas2019} for the
activation energy of the LeTID kinetics under 160~°C prolonged dark
anneals.

It is highly desirable to understand what happens if we change the
acceptor species. Are other group-III impurities able to interact
with H$_{2}$ molecules? Could the molecule dissociate upon reaction
with the acceptors? What are the consequences to the LeTID effect?
According to Ref.~\citep{Acker2022}, GaH pairs were effectively
formed in gallium-doped floating-zone Si structures that were annealed
in the dark. From Arrhenius plots, an activation energy of $1.04\pm0.01$~eV
was extracted for the growth rate of the GaH pair concentration, suggesting
a faster H$_{2}$ dissociation upon reaction with Ga than B.

Grant \emph{et~al.} \citep{Grant2020} compared LeTID of B- and Ga-based
cells, the former being subject to a regeneration treatment (see above)
at the last stage of fabrication. They found that under 1~sun illumination
at 75~°C, the B-based cells remained stable, whereas Ga-doped devices
showed a slight degradation. It was however found that 30-min dark
anneals in the range 200-300~°C resulted in suppression of the stabilizing
effect of the regeneration treatment, leading to strong LeTID in the
B-doped cells. The extent of LeTID in Ga-doped cells also increased
after the 200-300~°C anneals.

According to Ref.~\citep{Kwapil2021}, the degradation rate in Ga-doped
cells was different to what is usually observed in B-based devices.
In the Ga-doped structures, whereas almost no degradation occurred
under 1~sun illumination at 75~°C, a lifetime decrease was found
upon exposure to low light intensity. It was suggested that the dopant
species plays an active role in the reactions governing LeTID, possibly
involving differences in the properties of the respective acceptor-hydrogen
pairs.

Regarding the atomistic and electronic structure of the BH pairs,
the common view is that they form a three-center B-H-Si structure,
where H sits at the bond center site between B and its Si first neighbor.
The pairs are understood to be electrically inert, although during
the early stages of research in this field, intense discussions took
place on whether their neutralization was due to compensation or passivation
\citep{Pankove1985,Johnson1985,Pantelides1987,Stutzmann1987,Chang1988,Denteneer1989}.
In the first case, hole transfer takes place from B to H, driving
the formation of close $\textrm{B}^{-}\textrm{H}^{+}$ ionic pairs.
In the second case, H is covalently bound to Si, leaving B three-fold
coordinated and both B and H atoms are neutral. Experiments favor
the latter view, and it is now generally accepted that the hydrogen
truly passivates the acceptor.

Multiple H trapping by boron was also anticipated earlier \citep{Korpas1992,Borenstein1993}.
Based on junction capacitance measurements and first-principles calculations,
we recently proposed that a boron-dihydrogen complex, possessing a
donor level at 0.175 eV below the conduction band bottom, and showing
a large capture cross-section for electrons, could be responsible
for LeTID behavior based on boron-doped Si materials \citep{Guzman2021}.
This proposal raises many questions, starting with the formation/activation
and destruction/inhibition mechanisms of such complexes. And what
about dihydrogenation of other acceptors species? What are the properties
of the corresponding complexes?

Besides the above, additional fundamental questions regarding the
interaction of H with p-type dopants, have been left unanswered for
decades. These need to be addressed as well. We refer for instance,
to the unexpected trend observed for the dissociation energies of
$X$H pairs ($X$ being one of B, Al, Ga and In), which do not show
a monotonic variation with the acceptor atomic number \citep{Zundel1989}.
Also surprisingly, BH shows the lowest dissociation energy (compared
to other $X$H pairs). Being the smallest atom among the group-III
acceptors, boron is expected to allow for relatively more room to
accommodate H within the $X$-H-Si structure, and therefore to show
the largest binding energy to H.

Another puzzle worthy of mention is the observation of a splitting
pattern for the BH-related vibrational peak, measured by Raman spectroscopy
and positioned at 1903~cm$^{-1}$ under uniaxial stress along the
$\langle100\rangle$ direction \citep{Herrero1988}. Based on the
premise that this mode involves the stretching of a Si-H bond, it
was concluded that the B-H-Si structure could not be linear, and the
symmetry of the BH pair could not be trigonal. Instead, it was proposed
that the H atom should be located at an ``off-bond-centered” site
\citep{Herrero1988}. However, this clashes with the generality of
the theoretical results, where trigonal symmetry was found \citep{DeLeo1985,Bonapasta1987,Chang1988,Denteneer1989,Estreicher1989,Guzman2021}.

This work aims at providing a modern view of the solid-state physics
of acceptor-hydrogen interactions in p-type silicon. Effects of temperature
and the presence of minority carriers were given special attention.
Besides shedding light on the open issues identified above, we have
investigated the thermodynamics of the most probable solid-state reactions
involving hydrogen and group-III acceptors. Other competing reactions
involving hydrogen interaction with two of the most abundant impurities
in Si materials, namely oxygen and carbon, are also investigated.

We start by describing the methodologies in Sec.~\ref{sec:theory}.
In Sec.\ \ref{subsec:res:bonding} we analyze the acceptor-hydrogen
bonding chemistry, its relation to the passivation effect, as well
as the dynamics of H in the vicinity of the acceptors. In Sec.~\ref{subsec:res:form-diss}
we explore the formation and dissociation mechanism of acceptor-H
pairs in Si. Chemical trends, temperature and carrier trapping effects
are discussed. In Sec.~\ref{subsec:res:h2-xoc} we report on the
interactions of H$_{2}$ molecules with oxygen, carbon and group-III
acceptors in Si. Strain induced interactions, formation of remote
and closely spaced impurity-H$_{2}$ pairs, as well as properties
of the complexes resulting from intimate chemical reactions (which
lead to changes in the bonding properties of the reactants) are described.
The temperature-dependent energy balance of these reactions is also
reported at this stage. In Sec.~\ref{subsec:res:lvms} we look at
the local vibrational mode (LVM) frequencies of boron-hydrogen complexes
in Si. Here we describe a comparative study, with the results being
examined in the light of experimentally well characterized frequencies
from elemental boron and hydrogen defects. We end the paper in Sec.~\ref{sec:conclusions},
where we discuss our findings and the conclusions are drawn.

\section{Theoretical methods\label{sec:theory}}

\subsection{Semi-local and non-local all-electron energies\label{subsec:theory:energy}}

First-principles calculations were carried out using the density functional
Vienna Ab-initio Simulation Package (VASP) \citep{Kresse1993,Kresse1996a,Kresse1996b},
employing the projector-augmented wave (PAW) method for the treatment
of electronic core states \citep{Blochl1994}. A basis set of plane-waves
with kinetic energy of up to 400~eV was used to describe the Kohn-Sham
states. Total energies were evaluated self-consistently, using the
hybrid density functional of Heyd-Scuseria-Ernzerhof (HSE06) \citep{Heyd2003,Krukau2006}
with a numerical accuracy of $10^{-6}$~eV. When compared to generalized
gradient approximated (GGA) calculations \citep{Perdew1996}, which
underestimate the band gap of Si by nearly 50\%, the HSE06 functional
gives a band gap of 1.1~eV for Si. This is about the figure measured
experimentally, allowing us to obtain first-principles $\sim0.1$~eV-accurate
electronic transitions between gap states and the band edges.

Defect energies were found using 512-atom (defect-free) supercells
of Si, obtained by replication of $4\times4\times4$ conventional
unit cells (lattice constant $a_{0}=5.4318$~Å). Defect structures
were firstly optimized within the GGA approximation \citep{Perdew1996}
using $\mathbf{k}=\Gamma$ to sample the Brillouin zone (BZ), until
the largest force became lower than 0.01~eV/Å. Tests on selected
defects using a finer sampling ($\Gamma$-centered grid of $2\times2\times2$
points) were carried out in order to verify the quality of the forces
and structures. The total energies (electronic and ionic) of defects
were found from single-point HSE06-level calculations of the GGA-level
structures (keeping the band structure sampled at $\mathbf{k}=\Gamma$).

\subsection{Defect properties\label{subsec:theory:properties}}

Transition energy levels of defects, $E(q/q')$, quantify the location
of the Fermi level above/below which a defect is more stable in $q/q'$
charge state, respectively. These were evaluated with respect to the
valence band top ($E_{\textrm{v}}$) according to the usual methodology
(see for example Ref.~\citep{Coutinho2020} and references therein),

\begin{equation}
E(q/q')-E_{\textrm{v}}=-\frac{E(q)-E(q')}{q-q'}-E_{\textrm{v}}^{\textrm{KS}},\label{eq:levels}
\end{equation}
where $E(q)$ is the total energy of the defect, and $E_{\textrm{v}}^{\textrm{KS}}$
is the calculated valence band top (highest occupied Kohn-Sham state
at $\mathbf{k}=\Gamma$ of a 512-atom bulk supercell).

For non-zero charge states, the energies in Eq.~\ref{eq:levels}
are subject to a correction, $E(q)=\tilde{E}(q)+E_{\textrm{corr}}(q)$.
This aims at removing the spurious electrostatic interactions between
the artificial lattice of charges and neutralizing background that
are implicitly created in periodic calculations \citep{Freysoldt2009}.
Here $\tilde{E}(q)$ is the total energy as found for the periodic
system, while $E(q)$ is now the (approximate) energy of the aperiodic
problem. For a localized defect, $E_{\textrm{corr}}$ scales as $q^{2}/L$,
where $L$ is a characteristic length of the supercell. In the present
case we investigated single donors/acceptors ($q=\pm1$) and the corrections
were smaller than 0.1~eV.

It is known that periodic charge corrections tend to work better when
the charge trapped at the defect is well contained within the supercell
\citep{Komsa2012}. Some of the defects studied are shallow acceptors,
and for those we used an empirical approach, much in the spirit of
the marker method of Resende \emph{et~al.} \citep{Resende1999}.
This method relies on error cancellation and usually leads to sub-0.1~eV
error bars \citep{Coutinho2003}. Accordingly, $E_{\textrm{corr},X}(-1)$
for a shallow acceptor complex incorporating a group-III species $X$,
is obtained from the error of the calculated hole binding energy for
the respective isolated acceptor,

\begin{equation}
E_{\textrm{corr},X}(-1)=E_{\textrm{h},X}^{\textrm{exp}}-\left[\tilde{E}_{X}(-1)-\tilde{E}_{X}(0)-E_{\textrm{v}}^{\textrm{KS}}\right]\label{eq:corr}
\end{equation}
where $E_{\textrm{h},X}^{\textrm{exp}}=46$, 72, 74, 157~meV are
measured hole binding energies, for $X=\textrm{B}$, Al, Ga, and In
shallow acceptors in silicon, respectively \citep{Ramdas1981}, and
$\tilde{E}_{X}$ are (uncorrected) energies of supercells with the
appropriate substitutional acceptor. Values of $E_{\textrm{corr},X}$
were between $-20$ and $-13$~meV.

Vibrational modes and their frequencies were obtained within the harmonic
approximation using density functional perturbation theory (DFPT)
\citep{Baroni2001}. This method gives analytical access to the dynamical
matrix elements from the gradient of the electron density relative
to atomic displacement and, from here, to the vibrational properties
without having to displace the atoms explicitly. These calculations
were performed within the GGA-level, on supercells with 64 atoms,
a Brillouin zone sampling grid of $2\times2\times2$ special $\mathbf{k}$-points,
using $E_{\textrm{cut}}=500$~eV, and the relaxed structures had
residual forces below 0.005~eV/Å. Besides the calculation of local
vibrational modes, resonant mode frequencies were also found for the
calculation of vibrational free energies (see below).

The intensity of IR absorption bands associated with a LVM was estimated
from its oscillator strength \citep{Brueesch1986,Giannozzi1994},

\begin{equation}
f(m)=\sum_{i}\sum_{\alpha,j}\left|Z_{\alpha,ij}^{*}e_{\alpha,j}(m)\right|^{2}\label{eq:lvm-int}
\end{equation}
where the summation runs over all atoms with index $\alpha$, as well
as Cartesian coordinate indexes $i$ and $j$. The calculation of
the Born effective charge tensor for each atom, $Z_{\alpha,ij}^{*}=V\partial P_{i}/\partial R_{\alpha,j}$,
involves finding the macroscopic polarization ($\mathbf{P}$) of a
defective supercell with volume $V$, and its response to atomic displacements
($\mathbf{R}$). In practice this is done using a finite difference
method. The polarization was conveniently found within DFPT (see Sec.
II.D.2 of Ref.~\citep{Baroni2001} and Ref.~\citep{Gajdos2006}
for an implementation of the PAW method). Finally, the quantity $e_{\alpha,j}(m)$
is the $m$-th mode mass-weighted eigenvector.

The elastic coupling of defects to the lattice can be described by
the elastic dipole tensor ($P_{ij}$). Accordingly, the elastic energy
of a defective supercell subject to external homogeneous strain field
$\epsilon_{ij}$ is \citep{Bruneval2015,Wrobel2021},

\begin{equation}
E=Vc_{ijkl}\epsilon_{ij}\epsilon_{kl}-P_{ij}\epsilon_{ij},\label{eq:elenergy}
\end{equation}
where $c_{ijkl}$ are the elastic constants of the material. Minimization
of Eq.~\ref{eq:elenergy} with respect to strain leads to

\begin{equation}
P_{ij}=Vc_{ijkl}\epsilon_{kl}=-V\sigma_{ij}.\label{eq:eldipol}
\end{equation}

The first equality allows us to find the dipole tensor from the strain
$\epsilon_{kl}$ produced by a defect in a fully relaxed supercell
with optimized lattice vectors and atomic geometry. From the second
equality on the other hand, $P_{ij}$ can be obtained from the stress
$\sigma_{ij}$ developed across the supercell, whose volume $V$ and
shape are kept invariant upon relaxation of the atomistic structure.
The second approach is usually more practical, and it was the one
used by us.

The elastic dipole tensor is related to another useful quantity, namely
the relaxation volume tensor, $\Omega_{ij}=s_{ijkl}P_{kl}$, whose
trace is related to the elastic relaxation volume of the defect, $\Omega_{\textrm{rel}}=\textrm{Tr}\Omega$,
where $s_{ijkl}$ are the elastic compliance tensor elements of the
crystal \citep{Bruneval2015,Wrobel2021}. The relaxation volume quantifies
the macroscopic volume change of a sample upon defect introduction.

The bonding character of hydrogen to the dopants was investigated
using the Electron Localization Function (ELF) as defined by Becke
and Edgecombe \citep{Becke1990}, and later explored for bond topology
analysis by Silvi and Savin \citep{Silvi1994}. Accordingly,

\begin{equation}
\textrm{ELF}(\mathbf{r})=\frac{1}{1+\chi^{2}(\mathbf{r})},\label{eq:elf}
\end{equation}
where $\chi=D_{\textrm{P}}/D_{\textrm{h}}$ is the ratio of the kinetic
energy density due to Pauli repulsion between electron pairs to the
kinetic energy density of the homogeneous electron gas,

\begin{equation}
D_{\textrm{P}}(\mathbf{r})=\frac{1}{2}\sum_{i=\textrm{occ}}\left|\nabla\psi_{i}(\mathbf{r})\right|^{2}-\frac{1}{8}\frac{|\nabla n(\mathbf{r})|^{2}}{n(\mathbf{r})},\label{eq:elf-dp}
\end{equation}

\begin{equation}
D_{\textrm{h}}(\mathbf{r})=\frac{3}{10}\left(3\pi^{2}n(\mathbf{r})\right)^{5/3}.
\end{equation}
The second term in Eq.~\ref{eq:elf-dp} is the Boson-like kinetic
energy density of electrons \citep{Becke1990}, which is subtracted
to the Fermionic kinetic energy density (first term), so that $D_{\textrm{P}}$
becomes the contribution from Pauli repulsion effects alone. The electron
density $n$ is obtained from contributions of all occupied Kohn-Sham
orbitals, $n=\sum_{i=\textrm{occ}}|\psi_{i}|^{2}$. $D_{\textrm{P}}$
is therefore a positive definite quantity (the local Fermionic kinetic
energy is always greater than the Bosonic counterpart), and approaches
zero when electrons are alone or form pairs of opposite spins, i.e.,
when they are Boson-like. Hence, ELF provides a quantitative method
for mapping the localization of bonds and radicals.

The ELF is a scalar field with $0\leq\textrm{ELF}\leq1$, where 1
corresponds to a perfect localization either of a pair of opposite
spins (like in a covalent bond or a lone pair), or to an unpaired
electron (like in a radical). These correspond to maxima of ELF and
are referred to as \emph{attractors}. Values of ELF are small in the
border that separates highly localized regions, and $\textrm{ELF}=1/2$
when the electron density corresponds to that of the homogeneous electron
gas, which enters in the normalization of $\chi$ and provides a physical
reference. The ELF is represented graphically with help of isosurfaces
at a specific cut-off ELF$_{0}$, such that each enclosed volume defines
an ELF shell associated with a single attractor.

\subsection{Reaction barriers and free energies\label{subsec:theory:free}}

We have estimated the Helmholtz free energy of defect reactions. Besides
the (zero-temperature) potential energy change, this quantity describes
the change of electronic and ionic degrees of freedom between reactants
and products at finite temperatures, including the change in configurational
entropy ($\Delta F=\Delta F_{\textrm{elec}}+\Delta F_{\textrm{ion}}-T\Delta S_{\textrm{conf}}$).
The frozen core approximation of the pseudopotentials implies that
$\Delta F_{\textrm{elec}}$ accounts for the valence electrons (electron-electron
and electron-ion interactions), while analogous interactions involving
core electrons are accounted for by $\Delta F_{\textrm{ion}}$.

In many cases, we deal with electrically neutral defects or deep carrier
traps, for which the electronic entropy change vanishes or it is negligible
up to several hundred degrees Kelvin \citep{Estreicher2004}. Accordingly,
we assumed that $\Delta F_{\textrm{elec}}=\Delta E_{\textrm{elec}}$,
which is the static all-electron potential energy change. The ionic
free energy, $\Delta F_{\textrm{ion}}=\Delta E_{\textrm{ion}}+\Delta F_{\textrm{vib}}+\Delta F_{\textrm{rot}}$
accounts for the static ionic potential energy, as well as vibrational
and rotational free energies. For practical reasons, we reorganize
the above terms as

\begin{equation}
\Delta F=\Delta E+\Delta F_{\textrm{vib}}+\Delta F_{\textrm{rot}}-T\Delta S_{\textrm{conf}},\label{eq:free}
\end{equation}
where $\Delta E$ lumps together the stationary electronic and ionic
potential energies, here obtained from the density functional pseudopotential
calculations.

The Helmholtz free energy is calculated within the harmonic and dilute
approximations. The first assumes that the vibrational free energy
of reactants and products can be obtained from a set of $3N-3$ independent
harmonic oscillators with frequency $\omega_{i}$,

\begin{equation}
F_{\textrm{vib}}=k_{\textrm{B}}T\sum_{i=1}^{3N-3}\ln\left[2\sinh\left(\frac{\hbar\omega_{i}}{2k_{\textrm{B}}T}\right)\right],\label{eq:fvib}
\end{equation}
which already accounts for zero-point vibrational motion, where $k_{\textrm{B}}$
is the Boltzmann constant and $\hbar$ the reduced Plank constant.
The vibrational frequencies were obtained within the method already
described above. Regarding the dilute regime, it means that we ignore
defect-defect interactions, notably in the calculation of configurational
entropy.

The rotational free energy in Eq.~\ref{eq:free} is required for
reactions involving H$_{2}$ molecules. Isolated molecules at tetrahedral
interstitial sites are virtually free to rotate around their center
of mass \citep{Estreicher2001}. The free energy from this degree
of freedom is $\Delta F_{\textrm{rot}}=-k_{\textrm{B}}T\ln Z_{\textrm{rot}}$,
where the rotational partition function $Z_{\textrm{rot}}=Z_{\textrm{o}}^{g_{\textrm{o}}}Z_{\textrm{p}}^{g_{\textrm{p}}}$
accounts for independent populations of ortho (o) and para (p) spin-isomers
of the molecules,

\begin{equation}
Z_{\{\textrm{o,p}\}}=\sum_{j=\{\textrm{odd, even}\}}(2j+1)\exp\left[-j(j+1)\theta_{\textrm{rot}}/T\right],\label{eq:part}
\end{equation}
with natural occurring fractions $g_{\textrm{o}}=3/4$ and $g_{\textrm{p}}=1/4$.
The index $j$ runs over odd (even) integers for ortho (para) H$_{2}$,
and the characteristic rotational temperature is $\theta_{\textrm{rot}}=73$~K
\citep{Estreicher2004}.

Some defects were found to be shallow acceptors, and for reactions
involving those species, we have to add a term $\Delta F_{\textrm{elec,h}}$
to Eq.~\ref{eq:free}, which accounts for the electronic free energy
change due to thermal emission of holes across the reaction. We restrict
our analysis to temperatures and carrier concentrations in the context
of LeTID. Hence, the free energy change is assumed to simply reflect
the change in free carrier concentration due to full ionization/passivation
of the acceptors. Details for the estimation of $\Delta F_{\textrm{elec,h}}$
and configurational entropy are provided in Appendixes~\ref{apdx:a}-\ref{apdx:d}.

The above method has been successfully used for the study of thermodynamic
properties of defects in crystals \citep{AlMushadani2003,Murali2015,Zhang2018},
including hydrogen complexes in silicon \citep{Estreicher2004,Gomes2022}.
In the latter case, it was found that anharmonic effects become sizable
above $T\sim400\textrm{-}600$~K, in which case the conclusions drawn
must be considered as qualitative \citep{Gomes2022}.

The potential energy barriers of reactions were investigated using
the climbing image nudged elastic band method (NEB) \citep{Henkelman2000}.
These calculations were carried out using 64-atom supercells and $\Gamma$-centered
$2\times2\times2$ grids for sampling the BZ. First, we started by
setting up an array of up to 11 supercells linearly interpolated between
the stable end-configurations. On a second step, a NEB relaxation
was performed with forces calculated at the GGA level. In a final
and third step, the previous (GGA-grade) structures were used to obtain
the minimum energy path (MEP) by performing single-point calculations
at HSE06 level (keeping the $2\times2\times2$ BZ sampling grid).

\subsection{Notation and conventions\label{subsec:theory:notation}}

Before proceeding to the results section, we write a few words about
notation and conventions adopted. The reaction energy of a process
$A\rightarrow B$ is evaluated as $\Delta E_{\textrm{R}}=E_{B}-E_{A}$
and the corresponding binding energy (if applicable) is given by $E_{\textrm{b}}=-\Delta E_{\textrm{R}}$.
Accordingly, $E_{\textrm{b}}>0$ for an exothermic reaction, where
$E_{A}$ and $E_{B}$ are energies of states $A$ (reactants) and
$B$ (products), respectively.

Acceptor species $X=\textrm{B}$, Al, Ga, and In, are by default,
assumed to occupy a substitutional site. Neutral and negative states
are respectively referred to as $X^{0}$ and $X^{-}$. Carbon and
oxygen impurities are considered to occupy substitutional and interstitial
(bond centered) sites, respectively, and are referred to simply as
C and O defects. Should any ambiguity arise, they are identified with
commonly used subscripts as $X_{\textrm{s}}$ and $X_{\textrm{i}}$
, respectively.

Depending on its charge state, interstitial hydrogen in Si can adopt
different configurations \citep{Denteneer1989,VandeWalle1989}. By
default, we assume that isolated positive and negative species stand
for a bond-centered proton and anti-bonding hydride ion, respectively.
They are respectively represented by H$^{+}$ and H$^{-}$. The energy
of the often considered tetrahedral interstitial H$^{-}$ state was
only 0.04\,eV higher than the anti-bonding one. Such tiny difference
suggests that H$^{-}$ is delocalized over the tetrahedral cage and
effectively shows an \emph{averaged} tetrahedral symmetry. Neutral
hydrogen is most stable at the bond center site, and that is what
H$^{0}$ stands for (see Ref.~\citep{Gomes2022} and references therein
for a recent review of atomic and dimerized H in Si). Subscripted
H symbols like H$_{\textrm{AB}}$ and H$_{\textrm{BC}}$ (referring
to the impurity located at anti-bonding and bond-centered sites) are
avoided. This terminology is, however, necessary when addressing inequivalent
H atoms in a complex (e.g. H$_{2}^{*}$ and some boron-dihydride complexes).

As already mentioned, hydrogen in Si can form dimers, with at least
H$_{2}^{*}$ and H$_{2}$ molecules having been identified experimentally.
We will be referring to the molecules (located at the tetrahedral
interstitial sites) simply as H$_{2}$. Other H-related dimers, in
particular boron and carbon dihydride complexes have specific labels,
namely BH$_{2}$ and CH$_{2}$, respectively.

\section{Results\label{sec:results}}

\subsection{On the nature of the acceptor-H bonding\label{subsec:res:bonding}}

\begin{figure}
\includegraphics[width=7cm]{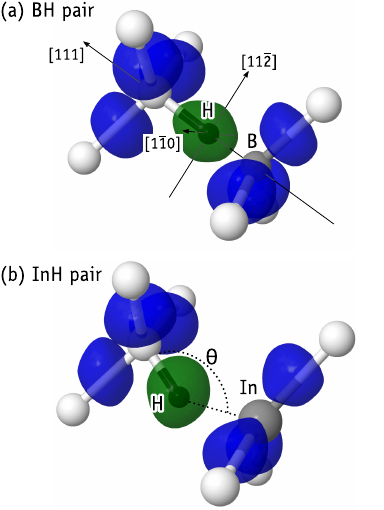}\caption{\label{fig1}Isosurface of the valence electron localization function
at $\textrm{ELF}=0.8$ for (a) BH and (b) InH complexes in silicon.
Blue surfaces represent ELF shells associated with Si-Si and Si-$X$
bonds. Green surfaces represent protonated ELF shells associated with
Si-H bonds. Silicon, hydrogen, and the acceptors are shown in white,
black and gray, respectively. The In-H-Si angle and crystallographic
directions are also indicated.}
\end{figure}

In agreement with previous reports \citep{Pankove1985,DeLeo1985,Bonapasta1987,Chang1988,Denteneer1989,Estreicher1989},
our results show that the H atom in $X$H pairs finds its most favorable
location between the acceptor and one of its Si first neighbors, forming
a three-atom $X$-H-Si structure. For the case of BH, the point group
symmetry of the resulting geometry is trigonal ($C_{3v}$), whereas
the pairs involving heavier acceptors show a bent $X$-H-Si unit,
forming an angle $\theta=137$°, 150°, and 138° for AlH, GaH and InH,
respectively. Equilibrium structures of BH and InH pairs are represented
in Figure~\ref{fig1}.

The nature of the local bonding of $X$H pairs was investigated with
help of the ELF. Figs.~\ref{fig1}(a) and \ref{fig1}(b) show ELF
isosurfaces for BH and InH pairs, respectively, using a cut-off $\textrm{ELF}=0.8$.
We find two types of ELF shells associated with covalent bonds between
pairs of atoms. The first kind are colored in blue and enclose ELF
attractors that lie between a pair of atoms (Si-Si and Si-$X$ bonds).
Their localization is typical of covalent homopolar bonds. The second
kind are protonated shells, clearly enclosing the H nuclei and displaced
toward the nearest Si atom (see Figure~2 of Ref.~\citep{Silvi1994}
for details of bond classification using ELF). All shells were found
to be populated with approximately 2 electrons. This result demonstrates
a genuine passivation effect via formation of a Si-H covalent bond
(not a compensation effect), and that the stable pairs should not
be viewed as adjacent ionic $\textrm{X}^{-}\textrm{H}^{+}$ moieties.
Instead, and as proposed by Pankove and co-workers \citep{Pankove1985},
they are best described as fully saturated $\equiv\textrm{Si-H}\cdots\textrm{X}\equiv$
structures, where Si, $X$ and H are four-fold, three-fold and mono
coordinated, respectively. The three dots between H and $X$ represent
a steric interaction that depends on the size of the $X$ species.

\noindent 

Of course, the calculated structure of BH is at variance with the
one proposed by Herrero and Stutzmann \citep{Herrero1988}, which
resulted from interpretation of the stress splitting of the BH-related
1903~cm$^{-1}$ Raman peak under $\langle100\rangle$ uniaxial stress.
We propose an alternative explanation, which reconciles the calculations
with the stress experiments of Ref.~\citep{Herrero1988} -- the
BH structure is trigonal, but the Raman peak at 1903~cm$^{-1}$ results
from the coupling of the H-Si stretching mode with an undetected low-frequency
doublet, possibly involving motion of H in the plane perpendicular
to the symmetry axis. In that case, the measurements would reflect
the splitting of the doublet under uniaxial stress.

The existence of a low-frequency degree of freedom coupled to the
stretching mode in the $X$H pairs was actually postulated earlier
by Stavola and co-workers \citep{Stavola1987,Stavola1988a} in order
to explain the observation of a pronounced red-shift of all acceptor-H
stretching band frequencies (by more than 30~cm$^{-1}$) upon raising
the temperature from liquid He up to room temperature \citep{Suezawa2002}.
Accordingly, with increasing the temperature, excited states of the
low-frequency mode become increasingly populated, offsetting the transition
energy related to the stretching mode by a few cm$^{-1}$. As noted
in Ref.~\citep{Stavola1987}, the above picture is reminiscent of
the vibrational spectra of interstitial oxygen, which also involves
a stretching mode of a three-center Si-O-Si unit coupled to a two-dimensional
low frequency mode of the order of 30~cm$^{-1}$, involving the motion
of oxygen in the $(111)$ plane \citep{Kaneta2003,Kaneta1990,Lassmann2012}.

We did not find a H-related wagging mode for BH above the Raman frequency
of Si. We did actually find several low frequency modes in the range
of 10-50~cm$^{-1}$ involving H motion in the $(111)$ plane. However,
because they are strongly mixed with the crystalline states, and the
magnitude of their frequency is as low as the error bar of the calculations,
all we can say is that (i) stress-splitting and temperature-dependent
optical absorption experiments point to the existence of such a low
frequency state, and that (ii) theory does not rule it out.

For $X$H complexes with heavier acceptors (showing off-bond-centered
structures), a low frequency motion could involve the rotation of
H around the bond center, leading to a roto-vibrational coupled model.
Indeed, from the energies of relaxed $X\cdots$H-Si structures with
H displaced along $[1\bar{1}0]$ and $[11\bar{2}]$, we estimate that
the warping of the potential energy path for H rotation around the
bond center site involves three equidistant barriers with height in
the range 5-20~meV only. This is smaller than typical zero-point
motion energies of H defects, suggesting that for the complexes with
heavier $X$ atoms ($X=\textrm{Al}$ , Ga, In) the H atom is delocalized
over a \emph{donut} around the BC site, effectively showing trigonal
symmetry like BH.

Another interesting observation against which $X$H models must be
tested, involves the hopping of H between equivalent bond centered
sites next to $X$. This was investigated by following the temperature-dependence
of the recovery of the stress induced dichroism in the absorption
spectra of samples where BH pairs were aligned under uniaxial stress
\citep{Stavola1988b}. From the data, a barrier of 0.19~eV was extracted.
This figure was already reproduced theoretically as $E_{\textrm{a}}=0.20$~eV
by Denteneer \emph{et~al.} \citep{Denteneer1989}. We found exactly
the same value using the NEB method.

Analogous experiments were also reported for the InH pair using perturbed
angular correlation spectroscopy to monitor the anisotropy of hyperfine
interactions of $^{111}$In ions paired with H \citep{Marx1996}.
The stress was applied \emph{in situ}, at room temperature, along
$\langle100\rangle$, $\langle110\rangle$ and $\langle111\rangle$
directions, and with magnitude up to 0.2~GPa (about the same that
was used for the alignment of BH in Ref.~\citep{Stavola1988b}).
According to the authors, no preferential alignment of the InH pairs
was observed. From a NEB calculation we found that the barrier for
the jump of H between In-Si bonds amounts to $E_{\textrm{a}}=0.71$~eV,
$i.e.$ considerably higher than that for H motion in BH. This could
therefore explain the lack of a stress-induced alignment of the InH
hyperfine signal at room temperature \citep{Marx1996}. Such high
barrier is attributed to the relatively large displacement of H in
the transition state, which approaches the tetrahedral interstitial
site next to In along $\langle100\rangle$. For the case of BH, the
transition state consists of a relatively small BH dimer aligned along
$\langle100\rangle$, sharing the substitutional site.

\subsection{Formation and dissociation mechanisms of acceptor-H pairs\label{subsec:res:form-diss}}

\subsubsection{Binding energy of correlated acceptor-H pairs}

\begin{figure}
\includegraphics[width=8.5cm]{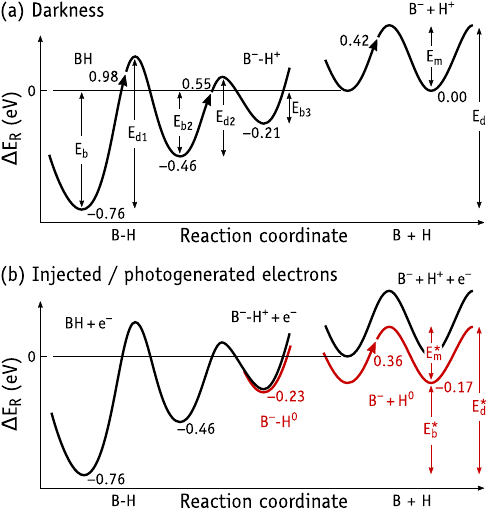}\caption{\label{fig2}Configuration coordinate diagrams with reaction energies
($\Delta E_{\textrm{R}}$) for boron-hydrogen pair formation in p-type
silicon (a) under equilibrium dark conditions, and (b) in red, showing
the effect of the presence of a free electron (\emph{e.g.} due to
current injection or photogeneration). Structures involving first
neighboring pairs (BH), second and third neighboring pairs (B-H),
and infinitely separated pairs ($\textrm{B}+\textrm{H}$) are considered.
The calculated migration barrier of hydrogen ($E_{\textrm{m}}$) was
taken from Ref.~\citep{Gomes2022}. The zero energy reference corresponds
to the $\textrm{B}^{-}+\textrm{H}^{+}$ state. Calculated figures
for other dopants are reported in Table~\ref{tab1}.}
\end{figure}

We now look at the reaction $X^{-}+\textrm{H}^{+}\rightarrow X\textrm{H}$.
This is an important route for formation of $X$H in hydrogenated
p-type Si. To find the binding energy of $X$H, the energy of the
reactants was calculated using (charge-corrected) independent supercells.
We also investigated the stability of close $X^{-}\textrm{-H}^{+}$
pairs, with H sitting on second and third neighboring bonds to $X^{-}$.
A reaction coordinate diagram is shown in Fig.~\ref{fig2} for the
specific case of BH formation, where the pairing reaction energy $\Delta E_{\textrm{R}}$
is represented as a function of the separation between $\textrm{B}^{-}$
and $\textrm{H}^{+}$. The zero energy corresponds to the state of
fully independent reactants. We start by discussing BH formation/dissociation,
and after that, we report the results and trends obtained for other
acceptors.

\noindent 

The BH pair ground state has a calculated binding energy $E_{\textrm{b}}=0.76$~eV.
This value nearly coincides with the experimental value of 0.75~eV
obtained recently by Voronkov and Falster \citep{Voronkov2017}, and
hopefully this clears some doubts regarding early calculations which
found rather different figures, $E_{\textrm{b}}=2.5$~eV \citep{Chang1988}
and $E_{\textrm{b}}=0.59$~eV \citep{Denteneer1989}. The first-neighboring
BH pair is markedly more stable than second and third neighboring
B-H pairs with $E_{\textrm{b2}}=0.46$~eV and $E_{\textrm{b3}}=0.21$~eV.
To check for the existence (or not) of a capture barrier along the
pairing reaction just before reaching the BH ground state structure,
we calculated the potential energy surface for the hydrogen jump from
first to second, and from second to third neighboring bond center
sites with respect to boron. These calculations were performed using
the NEB method spanning a total of 11 structure images along each
path. As shown in Fig.~\ref{fig2}(a), the “first dissociation” jump
involves overcoming a barrier of nearly $E_{\textrm{d1}}=1$~eV,
whereas the barrier for a second jump (to the third neighboring bond
center site) is only $E_{\textrm{d2}}=0.55$~eV high.

As $\textrm{H}^{+}$ jumps further away from $\textrm{B}^{-}$, the
heights of the corresponding barriers are assumed to progressively
decrease toward the migration barrier of isolated $\textrm{H}^{+}$.
Using identical methodologies to those employed in this work, we recently
reported the migration barrier of $\textrm{H}^{+}$ in Si as $E_{\textrm{m}}=0.42$~eV
\citep{Gomes2022}, implying that our estimate for the dissociation
barrier of BH is $E_{\textrm{d}}=1.18$~eV. This figure is only 0.1~eV
below the experimental result of Zundel and Weber \citep{Zundel1989}.
Besides reassuring the validity of the formation/dissociation model
presented above, this difference provides us with an estimate of the
magnitude of the error bar for energy values of the order of 0.1~eV
in our calculations.

Based on the calculated interactions between hydrogen and other group-III
species in Si, we can carry out an analogous analysis regarding the
formation and dissociation of $X$H pairs with $X=\textrm{Al}$, Ga,
and In. The results are summarized in Table~\ref{tab1} for all species.
The following aspects are worthy of note: (i) The calculated dissociation
barriers (and binding energies) depend weakly on the acceptor species,
$1.18\,\textrm{eV}\leq E_{\textrm{d}}\leq1.37\,\textrm{eV}$; (ii)
BH shows a relatively smaller binding energy, and consequently a lower
dissociation barrier; (iii) Among the three $X$H pairs with heavier
acceptors, GaH shows a slightly smaller binding energy. All these
features are also reflected in the measurements of Zundel and Weber
\citep{Zundel1989} ($c.f.$ experimental dissociation barriers reproduced
in Table~\ref{tab1}).

\noindent 
\begin{table}
\begin{ruledtabular}
\noindent \caption{\label{tab1}Binding energies ($E_{\textrm{b}}$) and dissociation
energies ($E_{\textrm{d}}$) of acceptor-hydrogen ($X$H) pairs in
silicon. Starred quantities refer to $X\textrm{H}+\textrm{e}^{-}\rightleftarrows X^{-}+\textrm{H}^{0}$
reaction energies involving the capture/emission (rightward/leftward
directions, respectively) of an electron by hydrogen at a remote location
from the acceptor. See Figure~\ref{fig2} for a graphical description
and definition of the energies. All values are in eV.}
\begin{tabular}{lcccc}
Acceptor & B & Al & Ga & In\tabularnewline
\hline 
$E_{\textrm{b}}$ & 0.76 & 0.95 & 0.92 & 0.95\tabularnewline
$E_{\textrm{b2}}$ & 0.46 & 0.32 & 0.34 & 0.28\tabularnewline
$E_{\textrm{b3}}$ & 0.21 & 0.22 & 0.23 & 0.17\tabularnewline
$E_{\textrm{b}}^{*}$ & 0.59 & 0.76 & 0.73 & 0.76\tabularnewline
$E_{\textrm{d}}^{*}$ & 0.98 & 1.12 & 1.09 & 1.12\tabularnewline
$E_{\textrm{d1}}$ & 0.98 & 0.74 & 0.74 & 0.67\tabularnewline
$E_{\textrm{d}}$ & 1.18 & 1.37 & 1.34 & 1.37\tabularnewline
$E_{\textrm{d,exp}}$ \citep{Zundel1989} & 1.28 & 1.44 & 1.40 & 1.42\tabularnewline
\end{tabular}
\end{ruledtabular}

\end{table}

\subsubsection{Unexpected dissociation energy trend}

At least two obvious questions arise at this point. One of them is
why the dissociation (and binding) energy of BH is distinctively smaller
than the same quantity for the other pairs? After all, a small B atom
is expected to share its site volume with H easier than larger acceptors
are. The second question looks for an explanation for the non-monotonic
trend in the observed and calculated dissociation energies, which
show a small, unexpected decrease for GaH.

Let us first look at the issue pertaining to the small binding energy
of BH. As we move downward along the group-III column of the periodic
table, the increasing size of the acceptors lead to the buildup of
compressive strain around the dopant. Calculated values of their relaxation
volume ($\Omega_{\textrm{rel}}$) are, respectively, $-20.6$, 1.6,
1.4 and 8.1~Å$^{3}$ for substitutional B$^{-}$, Al$^{-}$, Ga$^{-}$,
and In$^{-}$. As described in Section~\ref{subsec:res:bonding},
$\Omega_{\textrm{rel}}$ quantifies the macroscopic volume change
of a sample upon defect introduction. The results clearly reflect
the tensile character of boron and the compressive nature of the remaining
dopants. On the other hand, each isolated proton is estimated to add
a volume $\Omega_{\textrm{rel}}=13.5$~Å$^{3}$ to the Si host.

From the above, we may expect a proton to be more stable in the middle
of a B$^{-}$-Si bond than in between In$^{-}$ and Si. We note however,
that the binding energy of H$^{+}$ to $X^{-}$ depends on the energy
balance between both reactants and products, and additional arguments
beyond the relaxation volume of the acceptors must be considered.
For instance, while the B-H-Si structure is linear, other $X$-H-Si
structures with larger acceptors are bent, and some of the strain
working against the pairing is released.

Relaxation volumes of $X$H pairs are $-6.3$, 11.8, 11.8 and 18.6~Å$^{3}$
for $X=\textrm{B}$ , Al, Ga, and In, respectively. Considering the
volumes of the isolated species reported above, we find that the whole
relaxation volume change along the reaction $X^{-}+\textrm{H}^{+}\rightarrow X\textrm{H}$
is $\Delta\Omega_{\textrm{R}}=0.8$, $-3.2$, $-3.1$, and $-3.0$~Å$^{3}$,
indicating that after all, for the purpose of comparing the binding
energies of different $X$H pairs, strain effects are not that important.

On the other hand, from the perspective of the local chemical bonding,
the reaction

\begin{equation}
X^{-}\textrm{-Si}+\textrm{Si-H}^{\text{+}}\textrm{-Si}\rightleftarrows X\cdots\textrm{H-Si}+\textrm{Si-Si},
\end{equation}
shows that the bond energy balance boils down to the energy for formation
of a $X\cdots\textrm{H}$ steric interaction against that for breaking
a $X^{-}\textrm{-Si}$ bond, plus formation of a remote Si-Si bond
against breaking of a H$^{+}$-Si unit in isolated bond centered H$^{+}$.
A few assumptions allow us to estimate the dominant chemical contribution
to $E_{\textrm{b}}$ as a function of $X$. Firstly, Si-Si and H$^{+}$-Si
bond energies do not depend on $X$ and can be disregarded. Secondly,
and following the ELF analysis, which shows that H is covalently connected
to Si, the steric interaction between H and the group-III anion must
be relatively weak and should not differ much among different dopants
-- the calculated distance between $X$ and H varies from 1.3 Å for
boron to 1.9 Å for indium.

The above suggests that a sizable contribution to the variation of
$E_{\textrm{b}}$ with different $X$ should come from changes in
the energy of the $X^{-}$-Si bond that has to be broken in the reactants
side. According to thermochemical data \citep{Luo2007}, B-Si and
Al-Si bonds store respectively 317~kJ/mol and 247~kJ/mol. The contribution
of these bonds to the drop in the total energy upon formation of $X$H
is 70~kJ/mol ($\sim0.7$~eV/pair) larger for $X=\textrm{Al}$ than
B. Hence, the cost of breaking the relatively stronger B-Si bond could
explain the weaker binding energy of BH in comparison to other $X$H
pairs.

Regarding the second issue -- the off-trend dissociation energy of
GaH with respect to neighboring AlH and InH -- we may find an explanation
if we look at the electronic structure of group-III atoms. Unlike
the other acceptors, Ga is well known to show a “d-block contraction”
effect. This results from incomplete screening of the Ga nuclear charge
by its relatively diffuse d-electrons, which enhance the contraction
of the outer s and p shells. The radii at which the magnitude of the
Al(3s) and Ga(4s) wavefunctions is greatest, are respectively $r_{\textrm{max}}=1.11$~Å
and 1.05~Å \citep{Mann1968}, clearly reflect this effect. Likewise,
wave functions of Al(3p) and Ga(4p) states reach their maximum at
$r_{\textrm{max}}=1.42$~Å and 1.40~Å, respectively \citep{Mann1968}.

The above facts suggests that Ga-related bonds are slightly shorter
than Al-related ones, thus explaining the relatively flatter angle
obtained for the Ga-H-Si geometry. The d-block contraction of Ga also
explains the formation of shorter and probably stronger Ga-Si bonds
for substitutional Ga (calculated as 2.39~Å) in comparison to Al-Si
bonds in substitutional Al (calculated as 2.41~Å long), thus justifying
the small decrease of the binding energy of GaH with respect to that
of AlH.

\subsubsection{Electronic activity of correlated acceptor-H pairs}

The electronic activity of $X$H pairs was investigated according
to Eq.~\ref{eq:levels}. Considering the $\textrm{Si-H}\cdots X$
ground state structures, we did not find transition levels within
the gap for any of the pairs. There must be however a separation between
$X$ and H, beyond which passivation is lost and a compensation effect
takes over. At this point, the donor level of bond centered H and
the acceptor level of $X$ emerge from the crystalline density of
states into the band gap. We estimated this effect for the case of
B$^{-}$-H$^{+}$, by calculating the $(-/0)$ transition of second
and third neighboring pairs. During this transition $\textrm{B}^{-}$
is kept in the negative charge state, while hydrogen changes from
H$^{0}$ to H$^{+}$. The results show that despite possessing a four-fold
coordinated boron atom and a Si-H-Si unit, a second neighboring $\textrm{B}^{-}$-H$^{+}$
complex is still electrically inactive.

Conversely, third neighboring pairs have a very shallow electron trap,
which in the case of B-H is only about 20~meV below the conduction
band. This is a $(-/0)$ transition level of the pair, but if we consider
the Si-H-Si unit only, this level could be described as a $(0/+)$
transition of H (next to a B$^{-}$ ionized acceptor). For isolated
H (infinitely separated from B), the donor transition is calculated
at $E_{\textrm{c}}-0.17$~eV, matching the experimentally measured
value \citep{Nielsen1999}.

The above results, in particular the change of the electronic structure
of the pair with respect to that of the isolated components, show
that the association of H and $X$ cannot lead to a compensation effect.
Combined with the ELF analysis, they indicate that a passivation mechanism
takes place via chemical saturation of the $\textrm{Si-H}\cdots X$
structure. Compensation certainly takes place, but only between remote
H$^{+}$ and $X^{-}$ species, driving their approach via Coulomb
attraction.

\subsubsection{Dissociation enhancement upon electron capture}

The solid black line of Fig.~\ref{fig2}(b) replicates the potential
energy surface of Fig.~\ref{fig2}(a). However, besides describing
the energy of H$^{+}$ and $X^{-}$ species, the red line represents
the energy of the system upon electron trapping by hydrogen, $X^{-}+\textrm{H}^{+}+\textrm{e}^{-}\rightarrow X^{-}+\textrm{H}^{0}$.
With this, we intend to analyze the effect of the presence of a free
electron in the conduction band, possibly injected or photogenerated.

Following our discussion above, electron trapping is only possible
when H is located beyond the third neighboring bond center site with
respect to boron. Electron capture leads to a drop in the energy between
20~meV for close pairs and 0.17~eV for uncorrelated pairs. Hence,
some heat is necessary to promote BH separation before any interaction
with minority electrons takes place.

Theoretical studies indicate that the diffusivity of neutral hydrogen
in silicon is drastically higher than that of H$^{+}$ \citep{VandeWalle1989,Estreicher2012,Gomes2022}.
The enhancement stems from the minute barrier for motion of metastable
H$_{\textrm{AB}}^{0}$ along the hexagonal channels of the Si lattice
\citep{VandeWalle1989,Estreicher2012}. Recent calculations show that
once H$^{0}$ reaches the metastable anti-bonding configuration (overcoming
a barrier of $E_{\textrm{m}}^{*}=0.36$~eV), the H atom could travel
long distances across a potential landscape that is flatter than the
zero-phonon energy of the defect \citep{Gomes2022}. We note that
Fig.~\ref{fig2}(b) only shows the higher ($E_{\textrm{m}}^{*}=0.36$~eV)
migration barrier, where a jump between two consecutive minima can
represent a large traveling distance across the crystalline hexagonal
channels.

Recently, B-H and Ga-H pairs were investigated by Fourier-transform
infra-red spectroscopy in the context of LeTID \citep{Weiser2020}.
The idea was to monitor the concentration of BH and GaH pairs upon
illumination. It was found that the concentrations of the pairs was
reduced to 80\% of its starting value after low intensity (5~mW/cm$^{2}$)
illumination at room temperature for 96~h.

This picture is consistent with earlier experiments, where the presence
of minority carriers was shown to enhance the dissociation of BH pairs
in Si \citep{Seager1991,Zundel1991}. As suggested by Seager and Anderson
\citep{Seager1991}, when free electrons are available, initially
the dissociation proceeds with hydrogen as H$^{+}$. However, its
subsequent neutralization upon electron capture leads to a marked
acceleration of the dissociation process.

This interpretation has been challenged by Herring\emph{ et~al}.
\citep{Herring2001}, who did not find any appreciable light-induced
dissociation rate of BH, at least up to the temperature of $T=320$~K
(in Refs.~\citep{Seager1991,Zundel1991} the injection-enhanced dissociation
was observed at $T\gtrsim120\:^{\circ}$C). The authors of Ref.~\citep{Herring2001}
argued that the electron capture by H$^{+}$ next to boron followed
by pair dissociation was unlikely. This is because at room temperature
and above, the ionization rate of H$^{0}$ is orders of magnitude
faster than electron capture, leaving little chance for the short-lived
H$^{0}$ atom to escape from B. Fig.~\ref{fig2}(b) indeed suggests
that the early dissociation steps do not depend on the presence of
minority carriers. Some heat is always necessary to break the BH pair.
In agreement with Ref.~\citep{Herring2001}, we may conclude that
free electrons do not have an impact on the rate of the early reaction
$\textrm{BH}\rightarrow\textrm{B}^{-}\textrm{-}\textrm{H}^{+}$. However,
upon annealing and for sufficiently large B-H separations, if we consider
an exceptionally fast migration of H$^{0}$ along the hexagonal channels
of the Si lattice, the capture of minority electrons by H$^{+}$ could
decrease the concentration of BH by smearing the H atoms across the
lattice and slowing the recovery of the pairs.

\subsubsection{Finite-temperature and zero-point motion effects}

Density functional theory aims at finding the many-body electronic
ground state energy and respective electron density subject to a static
external potential at $T=0$~K. It is therefore reasonable to question
about the importance of finite temperature effects to the results,
for instance, to the calculated reaction energies. Another issue which
is worth discussing is the contribution of zero-point motion to the
calculated quantities, and by the way, what would be the impact of
substitution of hydrogen by deuterium -- after all, for practical
reasons, H properties are often determined by performing measurements
involving the heavier isotope. For instance, the detectivity of D
via secondary ion mass spectrometry is nearly two orders of magnitude
better than that of H \citep{Stevie2016}. It is therefore important
to understand how reaction energies and barriers depend on the hydrogen
mass.

Regarding quantum mechanical effects like tunneling motion of nuclei,
in particular of H, they only stand out (in comparison to thermally
activated processes) at temperatures that are too low to grant them
relevance at room temperature and above. We leave these effects out
of the scope of this work.

\noindent 
\begin{figure}
\includegraphics[width=8.5cm]{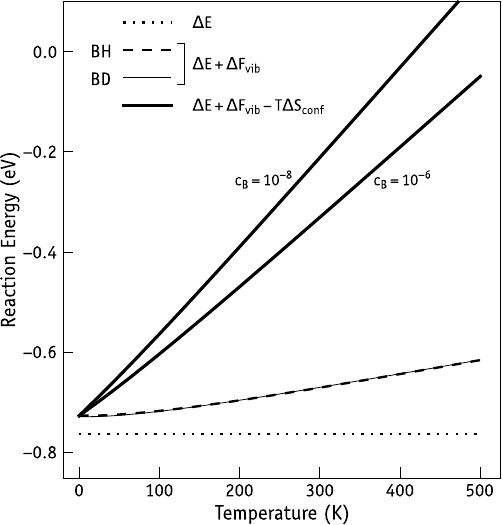}\caption{\label{fig3}Change in the Helmholtz free energy along the reaction
$\textrm{B}^{-}+\textrm{H}^{+}\rightarrow\textrm{BH}$ considering
the cumulative contributions from the all-electron-ion potential energy
(dotted), vibrational free energy (dashed) and configurational entropy
(thick solid). The vibrational free energy includes the zero-point
contribution. A plot of $\Delta E+\Delta F_{\textrm{vib}}$ for the
case of boron-deuterium pair formation (BD) is shown as a solid thin
line (almost undistinguishable from the dashed line of BH counterpart).
The total free energy change for two typical concentrations of boron
in Si are shown (thick solid lines).}
\end{figure}

Fig.~\ref{fig3} depicts static, dynamic and configurational contributions
to the Helmholtz free energy change across the reaction $\textrm{B}^{-}+\textrm{H}^{+}\rightarrow\textrm{BH}$.
The contributions are changes in (1) the all-electron-ion potential
energy ($\Delta E$) taken from many-body ground state energy calculations
within hybrid density functional theory; (2) the vibrational free
energy ($\Delta F_{\textrm{vib}}$), including zero-point motion ($\Delta E_{\textrm{ZP}}$),
obtained within the harmonic approximation; and also (3) the change
in configurational entropy across the reaction ($\Delta S_{\textrm{conf}}$).
The calculated free energy of a defect-free supercell had to be considered
in the products side of the reaction to comply with stoichiometric
balance.

We start by discussing the magnitude of the potential (static) terms.
The electron-ion energy change, $\Delta E=-0.76$~eV, is represented
by the horizontal dotted line and accounts for most of the reaction
energy. Zero-point energy due to atomic vibrations at $T=0$~K are
responsible for $\Delta E_{\textrm{ZP}}=37$~meV only, slightly decreasing
the binding energy to $E_{\textrm{b}}=0.72$~eV. The total potential
energy change ($\Delta E+\Delta E_{\textrm{ZP}}$) corresponds in
Fig.~\ref{fig3} to the value of the dependence shown by the dashed
line at $T=0$~K.

We now look at the effects of temperature and entropy across the $\textrm{B}^{-}+\textrm{H}^{+}\rightarrow\textrm{BH}$
reaction. The vibrational free energy, which describes the raise of
internal energy with $T$ due to vibrational excitations, the corresponding
increase in vibrational entropy, as well as zero point motion, is
shown in Fig.~\ref{fig3} as a dashed line ($\Delta E+\Delta F_{\textrm{vib}}$).
The changes in vibrational free energy are responsible for the temperature
dependence of the reaction free energy ($\Delta F_{\textrm{vib}}\sim0.12$~eV
upon increasing the temperature from 0~K to 500~K).

As for the configurational entropy change per BH defect across $\textrm{B}^{-}+\textrm{H}^{+}\rightarrow\textrm{BH}$,
we have (see Appendix~\ref{apdx:a}),

\begin{equation}
\Delta S_{\textrm{conf}}=k_{\textrm{B}}\ln(2c_{\textrm{B}}),
\end{equation}
where $c_{\textrm{B}}=n_{\textrm{B}}/n_{\textrm{Si}}$ is the fractional
concentration of boron, $n_{\textrm{B}}$ and $n_{\textrm{Si}}$ being
respectively the number of B dopants and Si sites in a crystalline
sample. As expected, the configurational entropy change across the
reaction is negative, thus favoring the reactants side with raising
the temperature. We are now in a position to estimate the temperature
above which BH is not stable anymore, and that is when $\Delta F=\Delta E+\Delta F_{\textrm{vib}}-T\Delta S_{\textrm{conf}}=0$.
Taking a reference doping concentration in the range $c_{\textrm{B}}=10^{-8}\textrm{-}10^{-6}$
(corresponding to $[\textrm{B}]=5\times10^{14}\textrm{-}5\times10^{16}$~cm$^{-3}$
and resistivity $\rho=27\textrm{-}0.35$~$\Omega$cm), we have $\Delta S_{\textrm{conf}}=-(1.52\textrm{-}1.13)$~meV/K.
The contribution of $-T\Delta S_{\textrm{conf}}$ to the free energy
change corresponds in Fig.~\ref{fig3} to the difference between
$\Delta F$ (solid lines) and $\Delta E+\Delta F_{\textrm{vib}}$
(dashed line). It can be seen from the figure that the reaction becomes
isothermic ($\Delta F=0$) at approximately $T\approx450$~K (or
about 180~°C). This temperature is just in the middle of the range,
140-220~°C, where the dissociation of BH pairs and recovery of electrical
activity of boron atoms occurred in float-zone-grown Si:B crystals
at equilibrium conditions (in the dark) \citep{Zundel1991}.

We also investigated the effect on the reaction energy of $\textrm{B}^{-}+\textrm{H}^{+}\rightarrow\textrm{BH}$
upon replacing hydrogen by deuterium. The electronic potential energy
and the configurational entropy of a defect population are not sensitive
to isotope change. Only zero-point motion effects and the vibrational
free energy depend on the mass of the nuclei. The change of the free
energy across $\textrm{B}^{-}+\textrm{D}^{+}\rightarrow\textrm{BD}$
is shown in Fig.~\ref{fig3} as a thin solid line, which is better
distinguishable from that related to BH formation (dashed line) close
to $T\sim0$~K, essentially reflecting small differences in zero-point
energy. The zero-point energy change for the deuterium reaction is
$\Delta E_{\textrm{ZP}}=0.34$~meV, only 3~meV smaller than the
same quantity for the hydrogen analogous reaction. Unaccounted anharmonic
effects, which are larger in H-related defects, could slightly increase
the difference.

We find two reasons that explain an almost identical temperature dependence
of H- and D-related reaction energies: (1) the high energy of hydrogen-related
vibrational frequencies; (2) the extreme localization of hydrogen-related
vibrational modes. Typical Si-H and Si-D vibrational frequencies in
the range $1500\textrm{-}2000$~cm$^{-1}$ correspond to energy quanta
of about 7-10 times $k_{\textrm{B}}T$ at room temperature. The first
excited state of these vibrations becomes populated above several
hundreds of Kelvin, and any meaningful \emph{mass-related difference}
in the contribution to vibrational entropy (or to the vibrational
free energy) is felt above thousands Kelvin only. On the other hand,
excited states of low energy modes are easily accessible and can play
a significant role in the reaction entropy change. These include strain-induced
modes which overlap many crystalline atoms. However, the virtually
identical strain field around H and D defects, combined with the extreme
localization of the H (and D) modes implies that these \emph{diffuse}
vibrations are effectively similar and decoupled from the H (and D)
atom.

\subsection{Interactions of H$_{2}$ molecules with acceptors, carbon and oxygen\label{subsec:res:h2-xoc}}

We now turn to the interactions of interstitial hydrogen molecules
with p-type dopants and two impurities commonly abundant in silicon
materials, namely substitutional carbon (C) and interstitial oxygen
(O). We start by discussing medium range interactions between correlated
$X$ and H$_{2}$ pairs separated by up to $\sim20$~Å. Then we will
look at short-range interactions, including molecular dissociation
next to the impurities.

\subsubsection{Strain interactions}

Hydrogen molecules are electrically inactive and are not subject to
long-range Coulomb attraction/repulsion by charged impurities. In
floating zone material, where the concentration of oxygen and carbon
are usually about $10^{16}$~cm$^{-3}$ or even less, H$_{2}$ molecules
represent most of the hydrogen stock available in samples exposed
to a hydrogen source at high-temperatures and subsequently quenched
\citep{Pritchard1998,Pritchard1999}. Isolated molecules are located
at tetrahedral interstitial sites of the Si lattice, and they become
mobile at temperatures slightly exceeding 300~K \citep{Markevich1998}.
In Cz-Si, most of the molecules are trapped near abundant oxygen impurities,
forming close O-H$_{2}$ pairs \citep{Pritchard1998}. Here, annealing
between room temperature and 160~°C leads to reversible displacement
of the equilibrium between the reactants and products sides of $\textrm{O}\textrm{-}\textrm{H}_{2}\rightleftarrows\textrm{O}+\textrm{H}_{2}$
\citep{Markevich1998}. As the molecules migrate across the silicon,
they are subject to the strain fields of several impurities and dopants
dissolved in the lattice, and that is what we detail in the following
paragraphs.

Interactions between H$_{2}$ and impurities/dopants were investigated
by placing H$_{2}$ on all symmetry-irreducible tetrahedral sites
of a 512-atom supercell with respect to a specific impurity/dopant.
Nearly 30 sites in total were identified.

\noindent 
\begin{figure}
\includegraphics[width=8.5cm]{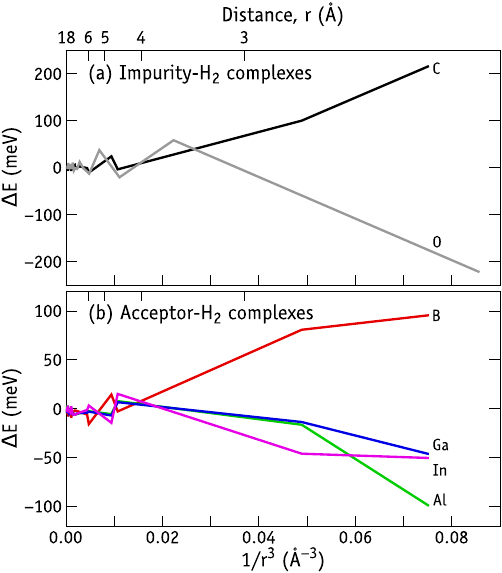}\caption{\label{fig4}Energy of (a) impurity-H$_{2}$ pairs and (b) acceptor-H$_{2}$
pairs in silicon as a function of the distance between the impurity/acceptor
and the molecule. Impurities considered are substitutional carbon
(C) and interstitial oxygen (O), both being electrically inactive.
The acceptors (B, Al, Ga, In) were all in the negative charge state.
The zero energy corresponds to the state of uncorrelated pairs.}
\end{figure}

Fig.~\ref{fig4} shows the relative energy of H$_{2}$-impurity/acceptor
pairs as a function of their separation, $r$ (upper horizontal axis).
The zero energy in the vertical axis is set to the state where the
molecule is infinitely separated from the dopant/impurity. The elastic
potential between defects decays as $r^{-3}$ \citep{Stoneham2001}
and that is how we define the scaling of the lower horizontal axis.
The scatter in the plots can be essentially attributed to the atomistic-range
interactions within the cubic host (not described by elasticity),
and to the anisotropy of interstitial oxygen. Based on its minute
barrier for rotation (few meV), effects from the elastic anisotropy
of the H$_{2}$ molecule are deemed very small.

Fig.~\ref{fig4}(a) depicts the results for H$_{2}$ interactions
with C and O. The first conclusion is that these two defects interact
rather differently with the molecules. While C is repulsive, O is
attractive. This could explain the observations in Cz-Si, where after
high-$T$ hydrogenation treatments with subsequent quenching, H$_{2}$
is trapped next to oxygen impurities. Our interpretation of these
results agrees with early calculations of Hourahine et al. \citep{Hourahine1997},
where it was found that bond centered O expands the volume of the
six equivalent nearest tetrahedral interstitial sites, making them
preferential for the molecule. Conversely, the short C-Si bonds create
a tensile elastic field around carbon, tightening the volume of close
interstitial sites, thus raising the energy of the resulting complexes.

Another interesting conclusion is that the interaction energy decreases
very quickly, converging toward below $k_{\textrm{B}}T$ at room temperature
for distances of $r\gtrsim5$~Å only. This is the magnitude of a
typical capture radius for defect reactions in the absence of electrostatic
interactions.

The calculated binding energy $E_{\textrm{b}}=0.22$~eV for O-H$_{2}$
is in good agreement with the experiments of Markevich and Suesawa
\citep{Markevich1998}, who derived 0.28~eV for this quantity. Accordingly,
H$_{2}$ molecules were formed after quenching O-rich Si that was
previously put in contact with hydrogen gas at 1200~ºC. The molecules
were found to be mobile at about room temperature and became trapped
next to O impurities. At slightly higher temperatures, $T\gtrsim50$~ºC,
the O-H$_{2}$ complexes dissociated, making the molecules available
to participate in other reactions \citep{Markevich1998}.

Weak interactions were also found for acceptor-H$_{2}$ pairs. Again,
the tensile species (B) is repulsive for the molecules, whereas compressive
elements (Al, Ga, In) have larger $X$-Si bonds and increase the open
volume of their six equivalent first neighboring interstitial sites.
The binding energy is rather week ($<0.1$~eV) and the trend obtained
may also reflect the electronic radius of each individual species.
Interestingly, the strongest repulsive effect of carbon could suggest
a preferential interaction between H$_{2}$ and boron, even when comparable
quantities of both impurities are present \citep{Pritchard1999}.

\subsubsection{Dissociation of H$_{2}$ molecules next to dopants and carbon\label{subsec:diss-h2}}

We recently estimated that, in the absence of a catalyst for H$_{2}$
dissociation, the molecules should survive up to about 400~°C in
pristine Si \citep{Gomes2022}, well above the annealing temperature
of H$_{2}^{*}$ dimers ($\sim200$~°C) \citep{Holbech1993}. The
dissociation of the molecule in a crystalline region of the Si was
suggested to occur upon collision of H$_{2}$ with a Si-Si bond, possibly
leading to the transient formation of metastable H$_{2}^{*}$ before
dissociation. The activation potential energy for the process was
evaluated (using the same methodology employed in this work) as $E_{\textrm{a}}=1.62$~eV.

\begin{figure}
\includegraphics[width=8.5cm]{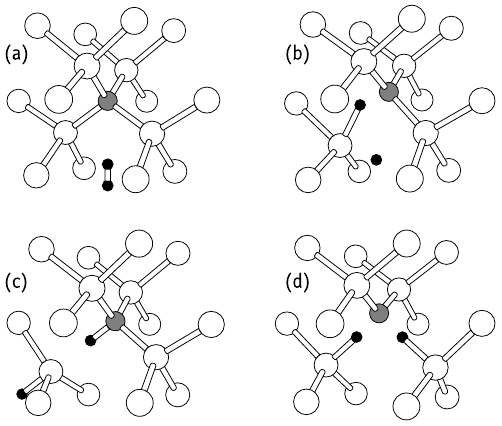}

\caption{\label{fig5}Mechanism for the dissociation of a H$_{2}$ molecule
next to a boron impurity and stable BH$_{2}$ complexes in silicon.
(a) Initial configuration with H$_{2}$ at the tetrahedral interstitial
site next to boron. (b) Metastable configuration after overcoming
the saddle point for dissociation. (c) Stable acceptor state of BH$_{2}$.
The most stable CH$_{2}$ complex is analogous, with boron being replaced
by carbon. (d) Stable donor state of BH$_{2}$. Si, B and H are shown
in white, gray, and black, respectively.}
\end{figure}

We investigated a similar reaction, but next to a boron impurity.
Likewise, the reaction between H$_{2}$ and substitutional carbon
was investigated for the sake of comparison. Carbon is well known
to interact with hydrogen and form stable CH$_{2}$ defects. Therefore,
it can compete with boron for the capture of H$_{2}$. The CH$_{2}$
complexes are electrically inert and stable up to $T\sim250$~°C
\citep{Markevich2001}. According to previous theoretical work \citep{Leary1998,McAfee2003,Estreicher2012},
the most stable form of the defect is illustrated in Fig.~\ref{fig5}(c).
The structure comprises a pair of $\equiv$C-H and $\equiv$Si-H units
aligned along a common trigonal axis, where the H atom connected to
carbon lies close to the center of a broken C-Si bond, while the H
atom connected to Si is at the anti-bonding site to the same broken
C-Si bond.

\noindent 

We looked at two reaction sites for H$_{2}$ next to boron. They involved
the overlap of the molecule either with first or second neighboring
bonds to the B atom (B-Si or Si-Si bonds, respectively). The mechanism
is almost identical in both cases, and they can fork into reactions
such as (1) $\textrm{H}_{2}+\textrm{B}^{-}\rightarrow\textrm{BH}_{2}^{-}$;
(2) $\textrm{H}_{2}+\textrm{B}^{-}+2\textrm{h}^{+}\rightarrow\textrm{BH}_{2}^{+}$;
or (3) $\textrm{H}_{2}+2\textrm{B}^{-}+2\textrm{h}^{+}\rightarrow2\textrm{BH}$.
In (1) and (2) we are implicitly anticipating that we found stable
negatively and positively charged BH$_{2}$ states. So far, the acceptor
state has been unreported. The donor state, on the other hand, was
recently identified by de~Guzman \emph{et~al.} \citep{Guzman2021}.

We found that the barrier along the minimum energy paths for the above
reactions involves the dissociation of H$_{2}$ next to first neighboring
B-Si bonds. Figs.~\ref{fig5}(a) and \ref{fig5}(b) help us to visualize
the early stages of the dissociation mechanism. They depict the structures
respectively before and after the transition state. The initial state,
comprising a molecule at the tetrahedral interstitial site next to
boron is shown in Fig.~\ref{fig5}(a). The molecule then moves toward
the B-Si bond, attaining the state shown in Fig.~\ref{fig5}(b).
The latter is a local minimum of energy at 0.49~eV above the initial
state. The transition state between the two configurations is 1~eV
above the initial state. To find the actual dissociation barrier of
H$_{2}$ upon reaction with B$^{-}$, we have to consider the 0.1~eV
offset between a close B$^{-}$-H$_{2}$ pair and infinitely separated
$\textrm{B}^{-}+\textrm{H}_{2}$ (\emph{c.f.} Fig.~\ref{fig4}(b)).
The activation energy for H$_{2}$ dissociation next to B is therefore
$E_{\textrm{d}}=1.1$~eV, nearly 0.5~eV lower than the analogous
quantity in pristine silicon (without the assistance of the dopant).
Note that the calculated dissociation barrier is larger than the energy
barrier that H$_{2}$ needs to surmount in order to migrate and get
close to the B$^{-}$ ion (0.78~eV \citep{Markevich1998}).

The metastable state of Fig.~\ref{fig5}(b) comprises a complex made
of a neutral BH pair next to a H$^{-}$ anion located close to the
tetrahedral interstitial site. Let us refer to it as $(\textrm{BH-H}^{-})^{*}$.
From here, several fast processes can take place, including H$^{-}$
relocation through a close hexagonal ring to form BH$_{2}^{-}$ as
depicted in Fig.~\ref{fig5}(c). The calculated barrier for this
step is 0.50~eV high, and overcoming it leads to completion of reaction
(1) referred above. Other possibilities may follow from the capture
of holes by stable BH$_{2}^{-}$ or metastable $(\textrm{BH-H}^{-})^{*}$
states. This could lead either to formation of BH$_{2}^{\text{+}}$
(reaction 2), whose ground state geometry is depicted in Fig.~\ref{fig5}(d),
or to the escape of neutral hydrogen from $(\textrm{BH-H}^{0})^{*}$,
which can subsequently capture a further hole and react with another
B$^{-}$ impurity (reaction 3).

The kinetics of formation and dissociation of BH pairs was recently
investigated during dark annealing treatments of floating-zone B-doped
Si wafers, passivated with silicon nitride on both faces, and subject
to a firing treatment \citep{Winter2021}. Activation energies for
formation and dissociation of the pairs of 1.29~eV and 1.22~eV were
extracted from the data cast in Arrhenius plots \citep{Winter2021}.
These figures are close to our calculated dissociation barrier of
H$_{2}$ next to B$^{-}$, and dissociation energy of a BH pair, 1.1~eV
and 1.2~eV, respectively. We note that the theoretical barrier is
a high bound of the true barrier. An alternative mechanism with lower
transition state energy could have escaped our search. That would
explain the lower activation energy (0.84~eV) for the formation rate
of BH pairs (from reaction between H$_{2}$ and boron) that was measured
by Walter \emph{et. al.} \citep{Walter2022}.

In another recent experiment, Acker \emph{et~al.} \citep{Acker2022}
studied the H$_{2}$ dissociation dynamics, and subsequent formation
of BH and GaH pairs in floating-zone Si doped with boron and gallium,
respectively. Activation energies of $E_{\textrm{a}}=1.20\pm0.02$~eV
and $E_{\textrm{a}}=1.04\pm0.01$ eV were found for the formation
rate of BH and GaH pairs, respectively. The first figure is again
very close to the calculated B-assisted dissociation barrier of the
molecule ($E_{\textrm{d}}=1.1$~eV). We did not calculate the Ga-assisted
dissociation barrier of H$_{2}$. However, considering that (i) unlike
boron, gallium does not show a repulsive strain field for the reaction
with H$_{2}$ (the repulsive barrier is 0.1 eV for boron \emph{c.f.}
Fig.~\ref{fig4}), and that (ii) judging from the similar pre-exponential
factors measured for BH and GaH formation ($\sim4\times10^{-8}\,\textrm{s}^{-1}$)
\citep{Acker2022}, both pairs are likely to share the same formation
mechanism, the calculated barrier for dissociation of H$_{2}$ when
the molecule is already next to boron (1~eV), could represent a good
approximation for the analogous figure involving the formation of
GaH pairs.

Regarding the reaction between H$_{2}$ and substitutional carbon,
the barrier for molecular dissociation next to the C-Si bond was estimated
as $E_{\textrm{d}}=1.15$~eV. The mechanism for the $\textrm{C}+\textrm{H}_{2}\rightarrow\textrm{CH}_{2}$
reaction is analogous to that involving boron. However, the metastable
state attained after overcoming the dissociation barrier, consists
of a rather stable structure $(\textrm{C-H}\cdots\textrm{H-Si})^{*}$,
with energy $-0.7$~eV below the initial state. Here, a C-Si bond
is converted into C-H and H-Si bonds, resulting in an electrically
inert structure which cannot interact with holes. The geometry of
$(\textrm{C-H}\cdots\textrm{H-Si})^{*}$ is non-linear with C and
Si bond angles showing substantial deviations from the tetrahedral
sp$^{3}$ geometry. Again, adding to the above barrier the 0.2~eV
off-set between the energies of the neighboring C-H$_{2}$ pair and
uncorrelated $\textrm{C}+\textrm{H}_{2}$ (\emph{c.f.} Fig.~\ref{fig4}(a)),
for the reaction $\textrm{C}+\textrm{H}_{2}\rightarrow\textrm{CH}_{2}$
we end up with a carbon-assisted H$_{2}$ dissociation barrier of
$E_{\textrm{d}}=1.35$~eV.

CH$_{2}$ complexes have been observed and studied in the past \citep{Leary1998,Hourahine2001,McAfee2003,Peng2011,Estreicher2012}.
They quickly form during the fast cooling that is applied after soaking
C-rich silicon in H$_{2}$ gas at around 1200~ºC. However, we do
not really know if the complexes form upon sequential capture of H
ions by carbon, or if they result from direct reaction between C and
H$_{2}$ molecules. Certainly, an interesting observation is that
if we start from as-quenched H$_{2}$-soaked boron-doped Fz-Si, subsequent
annealing treatments lead to (partial) loss of H$_{2}$ molecules,
and a concomitant raise in the concentration of BH pairs (see Fig.~2
of Ref.~\citep{Pritchard1999}). However, in carbon rich Fz-Si there
seems to be no reactions between H$_{2}$ molecules and C upon similar
treatments (see Fig.~3 of Ref.~\citep{Hourahine2001}). In the latter
case the CH$_{2}$ complexes are already formed in as-quenched samples,
and annealing leads to a decrease in their concentration only. Further
details on interactions between H$_{2}$ and C will be dealt with
in a separate paper.

\noindent 
\begin{figure}
\includegraphics[width=8.5cm]{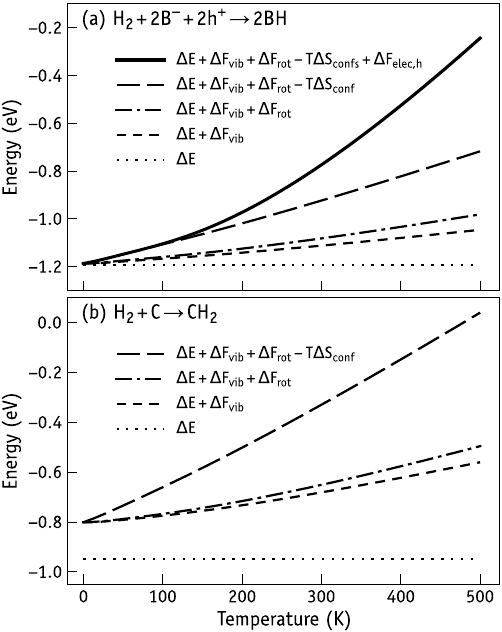}

\caption{\label{fig6}Cumulative contributions from the electron-ion potential
energy (dotted), vibrational free energy (dashed), rotational free
energy (dashed-dotted), configurational entropy (long dashed) and
hole free energy (solid) to the change in the Helmholtz free energy
($\Delta F$) along the reactions (a) $\textrm{H}_{2}+2\textrm{B}^{-}+2\textrm{h}^{+}\rightarrow2\textrm{BH}$
and (b) $\textrm{H}_{2}+\textrm{C}\rightarrow\textrm{CH}_{2}$. The
vibrational free energy accounts zero-point motion effects. Fractional
concentrations $c_{\textrm{B}}=c_{\textrm{C}}=10^{-6}$ and $c_{\textrm{H}}=10^{-8}$
were considered for the evaluation of configurational and electronic
entropy (see Appendix~\ref{apdx:d}).}
\end{figure}

It is useful to compare the most favorable reactions in p-type Si
involving H$_{2}$ with either boron or carbon. The reaction $\textrm{H}_{2}+2\textrm{B}^{-}+2\textrm{h}^{+}\rightarrow2\textrm{BH}$
is the one that maximizes the potential energy drop, corresponding
to $\Delta E_{\textrm{R}}=-1.19$~eV. Conversely, in carbon-rich
material H$_{2}$ is known to interact effectively with substitutional
carbon, and in that case, the reaction is $\textrm{H}_{2}+\textrm{C}\rightarrow\textrm{CH}_{2}$
with $\Delta E_{\textrm{R}}=-0.94$~eV, where CH$_{2}$ stands for
a neutral complex with a geometry identical to that shown in Fig.~\ref{fig5}(c).
Hence, from the perspective of the potential energy drop (at $T=0$~K),
both forms of hydrogen defects, namely BH pairs and CH$_{2}$ complexes,
show comparable stability, perhaps with a slight advantage to the
BH pairs.

Fig.~\ref{fig6} shows the cumulative contributions of several degrees
of freedom to the free energy change along reactions (a) $\textrm{H}_{2}+2\textrm{B}^{-}+2\textrm{h}^{+}\rightarrow2\textrm{BH}$
and (b) $\textrm{H}_{2}+\textrm{C}\rightarrow\textrm{CH}_{2}$. In
both cases, hydrogen is in the molecular form on the reactants side.
This implies that the energy reference of the products is the same.
The scaling of the axes was also made identical in both plots. At
first glance, we obviously note that the zero-point energy change
for reaction (a) is insignificant when compared to about 0.15~eV
in reaction (b). This is shown by the energy offset that separates
the electron-ion potential change $\Delta E$ (obtained within hybrid
density functional theory and represented as a dotted line) from the
other terms at $T=0$~K. This effect makes (b) less favorable, and
to great extent, is due to formation of stiffer and higher frequency
C-H and Si-H oscillators with wag modes in CH$_{2}$, against the
two softer Si-H oscillators with very low-energy wag modes in 2BH.
Another important aspect that favors BH formation in detriment of
CH$_{2}$ is the larger drop in configurational entropy in (b) than
in (a) (compare long dashed lines in Figs.~\ref{fig6}(a) and (b)),
so that $-T\Delta S_{\textrm{conf}}$ for BH pair formation grows
slower with increasing the temperature.

Reaction (a) involves the passivation of two shallow acceptors, and
above the carrier freezeout temperature of B-doped Si ($\sim80$~K),
it also involves the subtraction of two free holes per H$_{2}$ molecule.
We estimated the electronic free energy per hole ($\Delta F_{\textrm{elec,h}}$)
added to the electronic thermal bath of the sample. A change in the
free hole density $\Delta p=c_{\textrm{H}}[\textrm{Si}]=5\times10^{14}$~cm$^{-3}$
was considered, where $[\textrm{Si}]$ is the density of Si atoms
in crystalline Si in units of cm$^{-3}$, and that corresponds to
a full conversion of the available molecules ($c_{\textrm{H}}=10^{-8}$)
into BH pairs (the total boron available was $c_{\textrm{B}}=10^{-6}$).
Details of the calculation are presented in Appendix~\ref{apdx:d}.
The total free energy change (accounting for electronic, vibrational,
rotational, and configurational freedom) is represented in Figs.~\ref{fig6}(a)
as a solid thick line. It is clear that despite the rapid increase
in the free energy change with increasing the temperature, the free
energy drop for $\textrm{H}_{2}+2\textrm{B}^{-}+2\textrm{h}^{+}\rightarrow2\textrm{BH}$
is invariably larger than that for CH$_{2}$ formation shown in Fig.~\ref{fig6}(b).

We suggest that during the cooling of the Si, either after being subject
to a high temperature treatment in contact with hydrogen gas, or upon
firing a passivated Si structure covered by a H-rich passivating layer
(like in the fabrication of solar cells), H$_{2}$ molecules formed
within the bulk are more likely to react with boron than carbon, unless
the concentration of the latter is dominant.

The above conclusion is supported by both the smaller barrier for
the reaction of H$_{2}$ with B and the larger free energy drop per
molecule upon BH pair formation. It is also in line with the observed
H$_{2}$ reactions in different types of Si materials quenched from
high temperature treatments in H$_{2}$ gas. In cast multicrystalline
material, where carbon concentration is typically in the range $[\textrm{C}]\sim10^{17}\textrm{-}10^{18}$~cm$^{-3}$,
most hydrogen in as-quenched samples is found connected to carbon
in the form of CH$_{2}$ defects \citep{Peng2011}. On the other hand,
in boron doped floating zone Si with $[\textrm{C}]\lesssim[\textrm{B}]=10^{16}$~cm$^{-3}$,
as-cooled samples contained mostly isolated H$_{2}$ with 5\% of the
hydrogen already in the form of BH pairs \citep{Pritchard1999}. Annealing
at 160~ºC was sufficient for the molecules to travel and overcome
the reaction barrier (calculated here as 1.1~eV) before intimate
reaction with boron took place. A 1:1 correlation between the loss
of molecular hydrogen and formation of BH was found. Finally, in C-lean
n-type Cz-Si, the molecules are found trapped next to abundant O$_{\textrm{i}}$
impurities (where the binding energy of H$_{2}$ to oxygen was measured
as 0.26-0.28~eV) \citep{Pritchard1998,Markevich1998}. One puzzling
observation in the B-doped floating zone samples was that only about
50\% of molecular hydrogen was “consumed” during the conversion into
BH pairs. A saturation of the reaction was attained after about 24
hours at 160~ºC \citep{Pritchard1999}.

\subsubsection{Electronic structure of acceptor-H$_{2}$ complexes}

As briefly disclosed already, BH$_{2}$ can be both a donor and an
acceptor. The BH$_{2}^{+}$ donor state consists of two H atoms sitting
at the center of neighboring B-Si bonds. The atomistic geometry of
the donor, hereafter denoted as D is depicted in Fig.~\ref{fig5}(d)
and comprises two Si-H bonds next to undercoordinated boron. The one-electron
structure of the neutral state for this configuration shows a high-lying
highest occupied Kohn-Sham state edging the conduction band bottom.

Regarding the acceptor state, we found two stable BH$_{2}^{-}$ geometries
which resemble the H$_{2}^{*}$ complex. However, here the boron atom
replaces one of the two inequivalent Si atoms connected to H, leading
either to $\textrm{B-H}_{\textrm{BC}}\cdots\textrm{Si-H}_{\textrm{AB}}$
or $\textrm{Si-H}_{\textrm{BC}}\cdots\textrm{B-H}_{\textrm{AB}}$
configurations. They are respectively referred to as A and $\textrm{A}'$.
The first is depicted in Fig.~\ref{fig5}(c) and it is more stable
than $\textrm{A}'$ by 0.13~eV. In their neutral charge states, both
geometries lead to the appearance of a lowest unoccupied Kohn-Sham
state edging the valence band top, strongly indicating that they are
shallow acceptors.

In line with a recent report by us \citep{Guzman2021}, the donor
transition involving electron emission from neutral BH$_{2}^{0}(\textrm{D})$
to the conduction band, $\textrm{D}^{0}\rightarrow\textrm{D}^{+}+\textrm{e}^{-}$,
is calculated at $E_{\textrm{c}}-0.19$~eV. In that study we were
unaware of other configurations. We now found that A$^{0}$ is 0.43~eV
more stable than D$^{0}$, so that the actual donor transition should
be close to mid-gap, at $E(\textrm{A}^{0}/\textrm{D}^{+})-E_{\textrm{v}}=0.48$~eV
and involves two different structures.

\noindent 
\begin{table}
\begin{ruledtabular}
\noindent \caption{\label{tab2}Electronic transition levels and reaction energies of
$X\textrm{H}_{2}$ complexes in Si. The electronic transitions are
indicated graphically in Figure~\ref{fig7}. The energy of metastable
$\textrm{BH}_{2}^{-}(\textrm{A}')$ configuration with respect to
that of $\textrm{BH}_{2}^{-}(\textrm{A})$ is also shown.}
\begin{tabular}{lrrrr}
Acceptor ($X$) & B & Al & Ga & In\tabularnewline
\hline 
$E_{\textrm{c}}-E(\textrm{D}^{0}/\textrm{D}^{+})$ & 0.19 & $\sim0$ & $\sim0$ & $\sim0$\tabularnewline
$E(\textrm{A}^{0}/\textrm{D}^{+})-E_{\textrm{v}}$ & 0.48 & 1.20 & 1.15 & 1.29\tabularnewline
$E(\textrm{A}'^{0}/\textrm{D}^{+})-E_{\textrm{v}}$ & 0.61 & 1.38 & 1.49 & 1.80\tabularnewline
$E(\textrm{A}^{-}/\textrm{A}^{0})-E_{\textrm{v}}$ & 0.04 & 0.06 & 0.06 & 0.12\tabularnewline
$E(\textrm{A}'^{-}/\textrm{A}'^{0})-E_{\textrm{v}}$ & 0.04 & 0.08 & 0.07 & 0.17\tabularnewline
$E(\textrm{A}^{-}/\textrm{D}^{+})-E_{\textrm{v}}$ & 0.26 & 0.63 & 0.61 & 0.71\tabularnewline
$X\textrm{H}_{2}^{-}(\textrm{A})\rightarrow X\textrm{H}_{2}^{-}(\textrm{A}')$ & 0.13 & 0.20 & 0.35 & 0.55\tabularnewline
$2X\textrm{H}\rightarrow X_{\textrm{s}}^{-}+XH_{2}^{+}(\textrm{D})$ & 0.36 & 0.39 & 0.43 & 0.38\tabularnewline
$X\textrm{H}+\textrm{H}^{+}\rightarrow XH_{2}^{+}(\textrm{D})$ & $-$0.40 & $-$0.57 & $-$0.48 & $-$0.57\tabularnewline
$\textrm{H}_{2}+X_{\textrm{s}}^{-}\rightarrow X\textrm{H}_{2}^{-}(\textrm{A})$ & $-$0.31 & 0.07 & 0.14 & 0.22\tabularnewline
$X_{\textrm{s}}^{-}+\textrm{H}_{2}+2\textrm{h}^{+}\rightarrow X\textrm{H}_{2}^{+}(\textrm{D})$ & $-$0.83 & $-$1.19 & $-$1.07 & $-$1.19\tabularnewline
$2X_{\textrm{s}}^{-}+\textrm{H}_{2}+2\textrm{h}^{+}\rightarrow2X\textrm{H}$ & $-$1.19 & $-$1.58 & $-$1.50 & $-$1.58\tabularnewline
\end{tabular}
\end{ruledtabular}

\end{table}

As for the acceptor states (involving either structure A or $\textrm{A}'$),
we found them to be very shallow. A direct comparison of their electron
affinities with the same quantity for boron, leads us to the conclusion
that A and $\textrm{A}'$ defects are shallow acceptors with $(-/0)$
transitions very close to $E_{\textrm{v}}+40$~meV. We can understand
this result by noting that boron in A and $\textrm{A}'$ complexes
is four-fold coordinated.

Combining the shallow acceptor levels with the estimated location
of the donor transition, we find that BH$_{2}$ is a negative-$U$
complex with a metastable neutral state. The $(-/+)$ transition of
BH$_{2}$ is calculated at $E(\textrm{A}^{-}/\textrm{D}^{+})-E_{\textrm{v}}=0.26$~eV,
so that when the Fermi energy is below (or above) this level, the
complex is most likely to be found in the BH$_{2}^{+}(\textrm{D})$
(or BH$_{2}^{-}(\textrm{A})$) state \citep{Coutinho2020}.

\noindent 
\begin{figure*}
\includegraphics{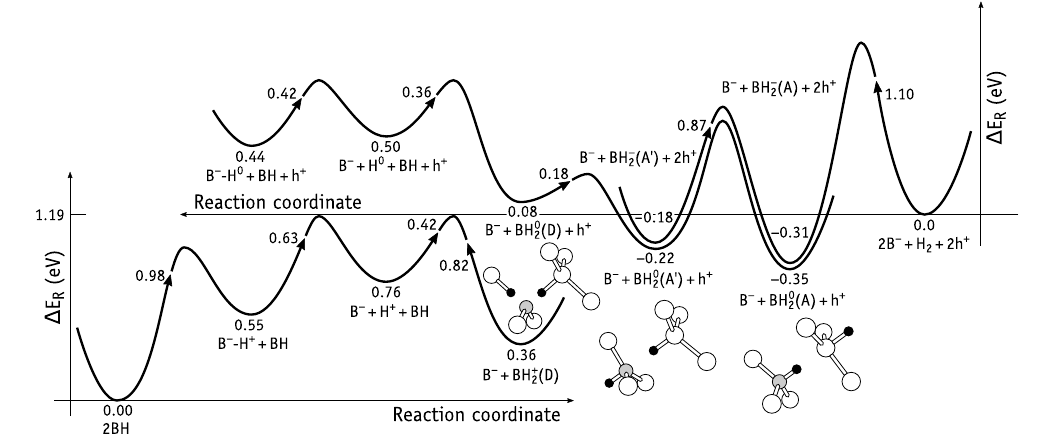}

\caption{\label{fig7}Configuration coordinate diagram of two substitutional
boron and two hydrogen atoms in silicon. Upper and lower diagrams
have zero energies for $2\textrm{B}^{-}+\textrm{H}_{2}+2\textrm{h}^{+}$
and 2BH states, respectively, which are separated by 1.19~eV. Energies
(in eV) of stable states are shown at the bottom of each local minimum.
Transition/reaction barriers are indicated next to arrow heads, and
they are relative to the nearest local minimum. Atomistic structures
of BH$_{2}$ complexes are presented next to their respective energy
minima. Si, B and H atoms are shown in white, gray and black, respectively.}
\end{figure*}

Dihydrogenated complexes with larger acceptors ($X\textrm{H}_{2}$
with $X=\textrm{Al}$, Ga, In) also show amphoteric and negative-$U$
character. For the donor state, the Si-H bonds notably deviate away
from $\langle111\rangle$ directions, implying that the bond angles
between the Si atoms (connected to H) and their ligands, display some
variations with respect to the perfect tetrahedral geometry. As for
the $X\textrm{H}_{2}^{-}(\textrm{A}')$ defects, they are also metastable
with respect to $X\textrm{H}_{2}^{-}(\textrm{A})$, respectively by
0.20, 0.35 and 0.55 eV for $X=\textrm{Al}$, Ga, and In. Notably,
two aspects distinguish the heavier $X\textrm{H}_{2}$ complexes from
BH$_{2}$. First, D$^{0}$ defects were found to ionize \emph{spontaneously},
with the electron on the highest occupied state being spread over
a conduction-band-like state. This is a strong indication of an effective-mass-like
donor defect. Second, the negative-$U$ transition levels for the
heavier $X\textrm{H}_{2}$ complexes were found in the upper part
of the gap, namely, $E(\textrm{A}^{-}/\textrm{D}^{+})-E_{\textrm{v}}=0.6$~eV
for AlH$_{2}$ and GaH$_{2}$, whereas $E(\textrm{A}^{-}/\textrm{D}^{+})-E_{\textrm{v}}=0.7$~eV
for InH$_{2}$. The calculated electronic transitions for all $X\textrm{H}_{2}$
complexes are summarized in Tab.~\ref{tab2}.

\subsubsection{Transformations and reconfigurations involving BH$_{2}$ complexes}

Considering the charge-dependent geometry of BH$_{2}$, it is instructive
to have a close look at possible transformations of this complex,
eventually involving the capture or emission of carriers. Fig.~\ref{fig7}
represents a configuration coordinate diagram of possible reactions
and transitions involving B and H in p-type silicon. The far right
and left of the CCD represent molecular hydrogen (away from boron)
and dissociated hydrogen molecules (paired with boron), respectively.
Between these two states, several reactions were considered, including
the formation of BH$_{2}$ complexes in the middle region. Energies
of local minima are represented below each potential basin along with
the respective chemical formula. Some of these figures were already
discussed above. The zero energy of the left (lower) part of the CCD
is the 2BH state. Conversely, the zero energy of the right (upper)
part of the diagram is the $2\textrm{B}^{-}+\textrm{H}_{2}+2\textrm{h}^{+}$
state. These are separated by 1.19~eV, representing the energy gain
per molecule after full passivation of boron. This result is close
to the measured activation energy $E_{\textrm{a}}=1.35\pm0.15$~eV
of the equilibrium constant of the reaction $\textrm{H}_{2}+2\textrm{B}^{-}+2\textrm{h}^{+}\rightleftarrows2\textrm{BH}$
in boron-doped Si \citep{Walter2022}. Fig.~\ref{fig7} also shows
several transformation barriers that were calculated from transition
state energies with respect to the basin of the corresponding initial
states. The energy barriers are indicated next to the arrow heads.

Much of the left part of the CCD of Fig.~\ref{fig7} has been described
in Sec.~\ref{subsec:res:form-diss} and detailed in Fig.~\ref{fig2}.
The dissociation energy of BH was estimated as 1.2~eV. However, under
above-band-gap illumination or carrier injection, the breaking of
BH into a close $\textrm{B}^{-}\textrm{-H}^{+}$ pair involves surmounting
a barrier of 0.98~eV, and when H$^{+}$ reaches the third neighboring
bond center site from boron and beyond (at least 0.55~eV above the
left-reference state), photogenerated or injected electrons can be
trapped by H$^{+}$, facilitating the escape of H$^{0}$ from the
ionic Coulomb field of B$^{-}$. Also, as pointed out already, during
individual jumps, H$^{0}$ atoms attain a metastable state in the
open interstitial regions of the lattice, where the potential energy
for motion is flatter than the zero-point energy. This implies that
this species can travel long distances athermally \citep{Gomes2022}.

The right end of the CCD of Fig.~\ref{fig7} describes the interaction
between the molecules and boron. Dissociation of H$_{2}$ assisted
by B involves surmounting a barrier of 1.1~eV, possibly leading to
formation of one BH pair plus release of H$^{+}$, which can subsequently
react with B$^{-}$. However, as we saw in Sec.~\ref{subsec:diss-h2},
it can also result in immediate formation of BH$_{2}^{-}(\textrm{A})$,
or BH$_{2}^{+}(\textrm{D})$ upon reconfiguration and capture of holes.
The potential energy barriers between BH$_{2}$ structures are relatively
small, especially the one separating BH$_{2}^{0}(\textrm{A}')$ from
BH$_{2}^{0}(\textrm{D})$. This amounts to 0.18~eV only, resulting
from the displacement of one of the H atoms in BH$_{2}^{0}(\textrm{D})$
to the anti-bonding site with respect to the B atom. The conversion
from BH$_{2}^{-}(\textrm{A}')$ into BH$_{2}^{-}(\textrm{A})$ starts
with the dissociation of the B-H bond in BH$_{2}^{-}(\textrm{A}')$,
followed by the displacement of the loose H$^{-}$ anion through hexagonal
sites to end up at the anti-bonding site next to the Si-H unit. The
first step of this mechanism involves surmounting a barrier estimated
as 0.87~eV. The barriers for subsequent jumps of H$^{-}$ are about
0.5~eV high \citep{Gomes2022}. Of course, at room temperature and
above, once the B-H bond is broken, H$^{-}$ is also likely capture
two holes to become H$^{+}$ at the bond center, escape from the neutral
BH pair, and finally be captured by B$^{-}$.

From our results we conclude that H$_{2}$ can dissociate upon interaction
with B$^{-}$, and that there is a strong thermodynamic drive for
dispersion of H in the form of BH pairs. Of course, BH$_{2}^{-}$
and BH$_{2}^{+}$ are possible intermediate species along the $2\textrm{B}^{-}+\textrm{H}_{2}+2\textrm{h}^{+}\rightarrow2\textrm{BH}$
reaction.

\noindent 
\begin{figure}
\includegraphics[width=8.5cm]{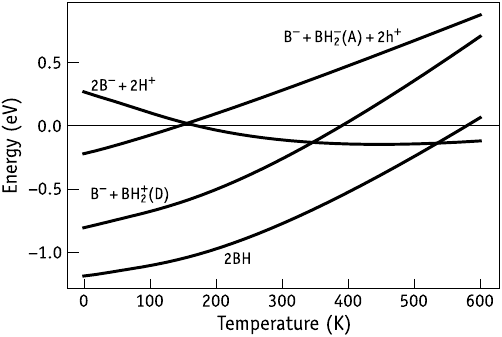}

\caption{\label{fig8}Temperature dependence of the free energy of some relevant
states involving a population of hydrogen and boron atoms in silicon,
with respect to the free energy of the $2\textrm{B}^{-}+\textrm{H}_{2}+2\textrm{h}^{+}$
state (represented by the horizontal line at zero energy). Configurational
and electronic entropy contributions were estimated by assuming concentrations
of substitutional B and interstitial H of $10^{-6}$ and $10^{-8}$,
respectively.}
\end{figure}

Fig.~\ref{fig8} provides a qualitative picture of the thermal stability
of several relevant states comprising a population of hydrogen and
boron atoms in silicon. Concentrations of substitutional B and interstitial
H were assumed to be $10^{-6}$ and $10^{-8}$, respectively. These
quantities enter in the evaluation of configurational and electronic
entropy of holes. The figure depicts the temperature dependence of
the free energy when all H is paired with B (2BH), when it is dissociated
from B and dispersed in the form of protons ($2\textrm{B}^{-}+2\textrm{H}^{+}$),
and when it forms boron-dihydrogen complexes ($\textrm{B}^{-}+\textrm{BH}_{2}^{+}$
and $\textrm{B}^{-}+\textrm{BH}_{2}^{-}+2\textrm{h}^{+}$). All free
energies are represented with respect to a common reference state
$2\textrm{B}^{-}+\textrm{H}_{2}+2\textrm{h}^{+}$ (shown as a horizontal
line).

In the analysis of Fig.~\ref{fig8} we should bear in mind several
approximations considered. Among these we highlight (i) the assumption
of a low concentration of impurities and the neglect of impurity-impurity
interactions; (ii) that H impurities and complexes are all in the
same form (for instance, in the 2BH state all hydrogen atoms are paired
with boron while unpaired boron atoms occupy substitutional positions);
(iii) limitations in the approximated treatment of configurational
and electronic entropy.

Fig.~\ref{fig8} shows that at $T=0$~K the 2BH state is 1.19~eV
below the $2\textrm{B}^{-}+\textrm{H}_{2}+2\textrm{h}^{+}$ reference.
This was already shown in Fig.~\ref{fig7}. With increasing temperature
BH becomes unstable with respect to dispersed H$^{+}$ at $T\apprge500$~K.
The main drive for the stabilization of the $2\textrm{B}^{-}+2\textrm{H}^{+}$
state is the dilution of hydrogen and the corresponding entropy increase.
It is noteworthy that the diagram cannot (and does not attempt to)
explain the formation of free molecules in hydrogenated Si quenched
from high temperatures -- the reference state ($2\textrm{B}^{-}+\textrm{H}_{2}+2\textrm{h}^{+}$)
is metastable with respect to $2\textrm{B}^{-}+2\textrm{H}^{+}$ above
150~K. On the other hand, H$_{2}$ formation is a non-equilibrium
process. As proposed in Ref.~\citep{Gomes2022}, the free energy
of formation of interstitial H$^{-}$ is reduced at high temperatures
and a small population of the anion is expected to show up. Combined
with a larger configurational entropy of the $\textrm{H}^{+}+\textrm{H}^{-}$
mix, a stabilization effect was proposed, allowing for the molecules
to form via Coulomb attraction between oppositely charged ions. This
effect deserves further investigation of its own.

Regarding the BH$_{2}$ complexes, they display a low thermal stability.
As shown in Fig.~\ref{fig8}, the relative free energy of the acceptor
state ($\textrm{B}^{-}+\textrm{BH}_{2}^{-}+2\textrm{h}^{+}$) becomes
positive well below room-temperature. This applies to p-type Si and
it means that the reaction of the molecule with B$^{-}$ (without
involvement of holes) is not favorable. If formed temporarily, BH$_{2}^{-}$
complexes are likely to dissociate or capture holes to end up in BH$_{2}^{+}$.
The difference between the $T$-dependence of the free energies of
$\textrm{B}^{-}+\textrm{BH}_{2}^{-}+2\textrm{h}^{+}$ and $\textrm{B}^{-}+\textrm{BH}_{2}^{+}$
states, mostly comes from the contribution of electronic entropy.
The donor state has less entropy due to capture of two holes. This
state is also metastable against 2BH across the whole temperature
range, and if formed, it has a barrier of only 0.82~eV preventing
its dissociation into more stable BH complexes (\emph{c.f.} Fig.~\ref{fig7}).
Hence, thermal dissociation of BH$_{2}^{+}$ and further formation
of BH pairs is likely to take place close to room temperature.

The BH$_{2}^{+}(\textrm{D})$ defect was recently suggested to be
responsible for an electron trap at $E_{\textrm{c}}-0.175$~eV (referred
to as E$_{0.175}$) in n-type Si co-doped with boron \citep{Guzman2021}.
The observation of a pronounced Poole-Frenkel effect suggested that
E$_{0.175}$ relates to a donor transition. Our results support that
conclusion -- we estimate the donor transition of BH$_{2}$(D) to
be located at 0.19~eV below $E_{\textrm{c}}$. Annealing experiments
have shown that E$_{0.175}$ starts to anneal out above 280~K under
open-circuit conditions \citep{Fattah2023}. This observation also
agrees with the finite temperature results in Fig.~\ref{fig8}, which
indicate that BH$_{2}^{+}$ should dissociate into $\textrm{B}^{-}+2\textrm{H}^{+}$
at about room temperature.

In n-type Si the lowest state of BH$_{2}$ is the shallow acceptor
BH$_{2}^{-}(\textrm{A})$ and not the donor state. However, in p-type
material the calculated thermal instability of BH$_{2}^{-}$ indicates
that this state cannot form (being less stable than $\textrm{B}^{-}+\textrm{H}_{2}$)
above $\sim100$~K. Hence, above this temperature, BH$_{2}$ should
be effectively viewed as donor defect (without an acceptor) for a
wide temperature range.

We finally point to the calculated potential energy changes pertaining
to the $2\textrm{B}^{-}+\textrm{H}_{2}+2\textrm{h}^{+}\rightarrow2\textrm{BH}$
reaction, including intermediate complexes, summarized at the bottom
of Tab.~\ref{tab2}. Analogous quantities are reported for the same
reactions involving heavier acceptors ($X=\textrm{Al}$, Ag, In).
Although we did not investigate reaction transition states for $X\neq\textrm{B}$,
based on the reaction energetics reported in Table~\ref{tab2} we
find that the CCD of Fig.~\ref{fig7} can be applied to the heavier
dopants as well, if we make a few adjustments. Several features/differences
are highlighted: (i) The $X^{-}+X\textrm{H}_{2}^{+}$ state is in
general metastable by approximately 0.4~eV with respect to the $2X\textrm{H}$
ground state; (ii) Unlike in B-doped material, formation of the shallow
acceptor $X\textrm{H}_{2}^{-}(\textrm{A})$ states (with $X=\textrm{Al}$,
Ga, In) upon reaction between H$_{2}$ and $X^{-}$ is endothermic
already at $T=0$~K; (iii) Like for the $X$H pairs, $X\textrm{H}_{2}^{+}$
has a smaller binding energy in B-doped Si, either with respect to
the release of molecules ($X^{-}+\textrm{H}_{2}+2\textrm{h}^{+}$),
or to the release of protons ($X\textrm{H}+\textrm{H}^{+}$); and
finally (iv) whereas BH$_{2}^{+}$ is a deep electron trap, other
$X\textrm{H}_{2}^{+}$ complexes are effective-mass-like shallow donors.
Regarding this latter aspect, Ref.~\citep{Fattah2023} provides a
proposal for the non-radiative recombination mechanism involving the
BH$_{2}^{+}$ complex, and a comparison with heavier \{Al,Ga,In\}H$_{2}^{+}$
complexes.

\subsection{Local vibrational modes of boron-hydrogen complexes\label{subsec:res:lvms}}

In this section, we explore the LVM frequencies of boron-hydrogen
complexes and their connection to the observations. The frequencies
are summarized in Tab.~\ref{tab3} and they are accompanied by their
calculated oscillator strengths. These can be related to the relative
amplitude of the corresponding infra-red absorption peaks. Due to
the harmonic approximation, the calculated frequencies tend to be
overestimated \citep{Hehre1986}. To facilitate the comparison between
theory and measurements, computed frequencies are often shifted or
scaled, typically by a factor $\lambda\gtrsim0.9$ \citep{Merrick2007}.
This effect is paradigmatic for Si-H stretching oscillations, whose
frequencies are often overestimated by more than 100~cm$^{-1}$ (see
Tab.~1 of Ref.~\citep{Nielsen1995} which reports values of Si-H
related modes of vacancy-hydrogen complexes in Si).

On the other hand, the quality of the description of the electron-electron
interactions for the calculation of the inter-atomic dynamical matrix
is also important. Here we use the GGA, which is known for underestimating
bond strengths, and consequently to soften the frequencies. The overall
error with respect to the measurements depends mostly on the relative
weight of the two opposite effects. Hence, a practical and often used
approach for better locating the frequencies is to calculate the frequency
deviation with respect to an experimentally well-characterized defect
mode (for instance a Si-H vibration in a VOH complex), and expect
a similar deviation for analogous modes in other defects (for instance
a Si-H vibration in VH).

The errors in the calculations can also show variations even when
dealing with defects that solely differ on the mass of one of its
elements (\emph{e.g.} isotope substitution). A bond-centered proton
has a calculated frequency for the asymmetric stretching mode of the
Si-H$^{+}$-Si unit of 2082~cm$^{-1}$. Experimentally, this was
attributed to a peak at 1998~cm$^{-1}$ in infra-red absorption spectrum
\citep{Budde2000}. Here the calculations overestimate the observations
by 84~cm$^{-1}$. On the other hand, in deuterated samples, the analogous
peak was observed at 1449~cm$^{-1}$ \citep{Budde2001} and we estimate
it at 1480~cm$^{-1}$ -- the underestimation is now 31~cm$^{-1}$
only, mostly due to the weaker anharmonicity of the heavier oscillator
(the description of the chemical bond in both cases is the same).
This effect is well known and abundantly discussed in the literature
(see for instance Refs.~\citep{Pruneda2002,Freysoldt2014}).

Previous theoretical work regarding the structure and vibrational
properties of substitutional boron and BH pairs is also well documented
(see for instance Refs.~\citep{DeLeo1985,Denteneer1989,Estreicher1989}).
We re-calculated the vibrational modes of these point defects to use
them as \emph{markers}, so that we can locate more accurately the
frequencies of modes from unexplored BH$_{2}$ complexes. For a review
on hydrogen and hydrogen-impurity properties in Si, including vibrational
properties, see Ref.~\citep{Estreicher2014b} and references therein.

A substitutional $^{11}$B species gives rise to a well-known vibrational
peak at 620~cm$^{-1}$, corresponding to a degenerate triplet state
involving vibrations of B against its Si neighbors along all three
$\langle100\rangle$ directions. We refer to these as B-Si bond stretching
modes and they are calculated at 621~cm$^{-1}$. Such an agreement
suggests that the anharmonicity is comparable to the softening effect
from the limitations in the exchange correlation description. Additionally,
and according to the above, calculations of comparable B-Si vibrations
from other defects should also display a similar vanishing error.
Tab.~\ref{tab3} clearly shows that the highest-frequency B-related
mode of the BH pair, calculated at 657~cm$^{-1}$ matches well the
experimental counterpart observed at 652~cm$^{-1}$\citep{Herrero1988}.
This is a doublet involving B oscillations in the plane perpendicular
to the $\textrm{Si-H}\cdots\textrm{B}$ axis, and results from the
$t_{2}\rightarrow e+a_{1}$ unfolding of the substitutional B mode,
due to the presence of H. The singlet state ($a_{1}$) is calculated
at 537~cm$^{-1}$, very close to 515~cm$^{-1}$, which was the calculated
highest frequency of a pristine silicon supercell.

\noindent 
\begin{table}
\begin{ruledtabular}
\noindent \caption{\label{tab3}Calculated local vibrational mode frequencies ($\nu_{\textrm{calc}}$
in cm$^{-1}$), their oscillator strengths ($f$ in $|e|$ units where
$e$ is the electron charge), symmetry representation within the respective
point group of symmetry, and bond localization, where \textquotedblleft str\textquotedblright{}
and \textquotedblleft wag\textquotedblright{} stand for stretching
and wagging modes, respectively. Symmetry point groups are $T_{d}$
(B$^{-}$), $D_{3d}$ (H$^{+}$, D$^{+}$), $C_{3v}$ (BH and BH$_{2}^{-}$)
and $C_{2v}$ (BH$_{2}^{+}$). Some experimental vibrational frequencies
($\nu_{\textrm{exp}}$ in cm$^{-1}$) and respective references are
also included. Calculated frequencies involving boron defects used
the mass of the most abundant $^{11}$B isotope.}
\begin{tabular}{rrrrrr}
Defect & $\nu_{\textrm{calc}}$ & $f$ & Sym. & $\nu_{\textrm{exp}}$ & Localization\tabularnewline
\hline 
H$^{+}$ & 2082 & 18.73 & $a_{2u}$ & 1998 \citep{Budde2000} & Si-H$^{+}$-Si (str)\tabularnewline
D$^{+}$ & 1480 & 12.72 & $a_{2u}$ & 1449 \citep{Budde2001} & Si-D$^{+}$-Si (str)\tabularnewline
\noalign{\vskip0.3cm}
B$^{-}$ & 621 & 0.22 & $t_{2}$ & 620 \citep{Herrero1988} & B-Si (str)\tabularnewline
\noalign{\vskip0.3cm}
BH & 2027 & 19.23 & $a_{1}$ & 1903 \citep{Stavola1988a} & Si-H$_{\textrm{BC}}\cdots$B (str)\tabularnewline
 & 657 & 0.07 & $e$ & 652 \citep{Herrero1988} & B-Si (str)\tabularnewline
 & 537 & 0.01 & $a_{1}$ &  & B-Si (str)\tabularnewline
\noalign{\vskip0.3cm}
BH$_{2}^{-}(\textrm{A})$ & 2265 & 3.15 & $a_{1}$ &  & B-H$_{\textrm{BC}}\cdots$Si (str)\tabularnewline
 & 1788 & 1.90 & $a_{1}$ &  & Si-H$_{\textrm{AB}}$ (str)\tabularnewline
 & 769 & 1.01 & $e$ &  & Si-H$_{\textrm{AB}}$ (wag)\tabularnewline
 & 707 & $\sim$0 & $e$ &  & B-H$_{\textrm{BC}}$ (wag)\tabularnewline
 & 604 & 0.10 & $e$ &  & B-Si (str)\tabularnewline
\noalign{\vskip0.3cm}
BH$_{2}^{-}(\textrm{A}')$ & 2175 & 5.45 & $a_{1}$ &  & B-H$_{\textrm{AB}}$ (str)\tabularnewline
 & 2087 & 1.10 & $a_{1}$ &  & Si-H$_{\textrm{BC}}\cdots$B (str)\tabularnewline
 & 860 & 0.95 & $e$ &  & B-H$_{\textrm{AB}}$ (wag)\tabularnewline
 & 680 & 0.20 & $e$ &  & B-Si (str)\tabularnewline
 & 561 & 0.52 & $e$ &  & Si-H$_{\textrm{BC}}$ (wag)\tabularnewline
\noalign{\vskip0.3cm}
BH$_{2}^{+}(\textrm{D})$ & 2186 & 13.98 & $a_{1}$ &  & Si-H$_{\textrm{BC}}\cdots$B (str)\tabularnewline
 & 2075 & 23.48 & $b_{1}$ &  & Si-H$_{\textrm{BC}}\cdots$B (str)\tabularnewline
 & 718 & 1.23 & $b_{2}$ &  & B-Si (str)\tabularnewline
 & 688 & 0.32 & $b_{1}$ &  & Si-H$_{\textrm{BC}}$ (wag)\tabularnewline
 & 544 & 2.43 & $a_{1}$ &  & Si-H$_{\textrm{BC}}$ (wag)\tabularnewline
\end{tabular}
\end{ruledtabular}

\end{table}

The BH complex also gives rise to a well-known stretching vibration
localized on the $\textrm{Si-H}\cdots\textrm{B}$ structure, leading
to an infra-red absorption band at 1903~cm$^{-1}$ \citep{Pankove1985,Johnson1985,Stavola1988a,Herrero1988,Weiser2020}).
Previous calculations slightly underestimated this frequency, pointing
to values in the range 1830-1880~cm$^{-1}$ \citep{DeLeo1985,Denteneer1989}.
We can only explain such nice agreement if we allow for some error
bars (\emph{e.g.} due to limitations in the description of the electronic
structure or boundary conditions) That frequency is here predicted
at 2027~cm$^{-1}$. It sets our overestimation to 124~cm$^{-1}$
for this type of mode, allowing us to better estimate similar modes
for BH$_{2}$ complexes.

All three structures of BH$_{2}$ (D, A and $\textrm{A}'$) give rise
to three stretching modes plus two wagging modes (Tab.~\ref{tab3}).
However, the modes are all non-degenerate for BH$_{2}^{+}$, in contrast
to both BH$_{2}^{-}$ complexes, where three of the modes are doubly
degenerate. BH$_{2}^{-}(\textrm{A})$ and BH$_{2}^{-}(\textrm{A}')$
complexes form $\textrm{Si-H}_{\textrm{BC}}\cdots\textrm{B-H}_{\textrm{AB}}$
and $\textrm{B-H}_{\textrm{BC}}\cdots\textrm{Si-H}_{\textrm{AB}}$
pairs, each having two H-related stretching vibrations, two H-related
wagging vibrations, and one B-related stretching vibration. The BH$_{2}^{+}(\textrm{D})$
complex also has two stretching modes and two wagging modes localized
on two equivalent Si-H units. Unlike for the acceptor states, where
Si-H and B-H oscillators are rather independent, H-modes in BH$_{2}^{+}(\textrm{D})$
are strongly coupled and form symmetric and asymmetry pairs.

Among the BH$_{2}$ complexes, BH$_{2}^{-}(\textrm{A}')$ and BH$_{2}^{+}(\textrm{D})$
possess modes localized on Si-H$_{\textrm{BC}}$ units, which are
of the same type of the BH-pair stretching mode. For the acceptor,
it has a frequency of 2087~cm$^{-1}$ and it is mostly localized
on the Si-H$_{\textrm{BC}}$ bond. Hence, after being subject to the
correction of 124~cm$^{-1}$, this mode is anticipated at 1963~cm$^{-1}$.
The donor state, on the other hand, has symmetric and anti-symmetric
modes on a $\textrm{Si-H}\cdots\textrm{B}\cdots\textrm{H-Si}$ structure.
They are calculated at 2186 and 2075~cm$^{-1}$, meaning that our
best estimate for these frequencies (after the correction) becomes
2062 and 1951~cm$^{-1}$, respectively.

So far, we could only find the work of Fukata \emph{et~al.} \citep{Fukata2005}
suggesting the formation of complexes with direct B-H bonds. The measurements
were performed in plasma-hydrogenated B-implanted n-type Si, where
several infra-red absorption bands in the range 2280-2470~cm$^{-1}$
were attributed to B-H vibrations. One of the bands could be related
to the most stable acceptor form of BH$_{2}$ with calculated B-H
stretching frequencies of 2265~cm$^{-1}$. Interestingly, the formation
of B-H related complexes/clusters was accompanied by an increase in
resistivity of the samples. Compensation by BH$_{2}^{-}(\textrm{A})$
complexes in the n-type samples could contribute to this effect.

From a theoretical perspective, we highlight that the two Si-H stretching
modes of BH$_{2}^{+}(\textrm{D})$ are predicted to have a strong
oscillator strength. Of course, a small concentration of such complexes
available in the Si may prevent their detection.

We end with a final note regarding the bonding of the BH pair. Analysis
of the isotope shifts can provide us with evidence that H is chemically
connected to Si (and not to B). Let us approximate the pair to an
oscillating $X$-H bond, $X$ being either B or Si, with frequency
$\nu$ and a spring constant that is independent of the masses $m_{X}$
and $m_{\textrm{Si}}$. If species $X$ is replaced by a lighter element
$X'$ with mass $m_{X'}$ the frequency is enhanced by $\Delta\nu=\nu(\sqrt{\mu/\mu'}-1)$,
where $\mu$ and $\mu'$ are reduced masses of $X$-H and $X'$-H
pairs, respectively. Now, if $X$ is a boron atom, for an oscillating
frequency of $\nu=1900$~cm$^{-1}$ we obtain $\Delta\nu\approx8$~cm$^{-1}$
upon replacing $X=^{11}\!\textrm{B}$ by $X'=^{10}\!\textrm{B}$ (boron
natural isotopes with respective abundance of 80\% and 20\%). Instead,
from the first-principles calculations we obtain a frequency shift
smaller than 2~cm$^{-1}$ for such substitution in the BH pair, suggesting
that the coupling between H and B atoms is much weaker than that for
a direct B-H bond. This result is supported by infra-red absorption
data, which show a minor shift in the BH band of nearly 1~cm$^{-1}$
after comparing floating-zone samples implanted either with $^{11}$B
or $^{10}$B ions \citep{Pajot1988,Watkins1990}.

If on the other hand, we assume that $X$ is a silicon atom, the harmonic
dimer model gives $\Delta\nu\approx2$~cm$^{-1}$ upon replacing
$X=^{30}\textrm{Si}$ by $X'=^{28}\textrm{Si}$ , and that matches
the analogous quantity that was calculated from first-principles for
the replacement of the Si atom in the Si-H$\cdots$B structure.

\section{Discussion and Conclusions\label{sec:conclusions}}

We presented a comprehensive set of first-principles calculations
regarding the properties of acceptor-hydrogen complexes in silicon,
including electronic transition energies, local vibrational modes
and respective intensities, defect-crystal strain coupling and electron
localization function. Reactions involving atomic and molecular hydrogen
with the acceptors were also investigated. From the calculated reaction
barriers, the change of free energy, and the charge state dependence
of these quantities, we provided a finite-temperature account of thermally-activated
and carrier-capture-activated processes involving acceptor-hydrogen
complexes in p-type Si.

The electronic structure of the BH pair was investigated as a function
of the distance between B and H units. Electric charges of infinitely
separated B$^{-}$ and H$^{+}$ pairs are mutually compensated. An
electron trap due to isolated H (donor transition) was estimated at
$E_{\textrm{c}}-0.19$~eV. As the B-H distance decreases, the H electron
trap and the B hole trap are increasingly affected by repulsive potentials,
and as a result, they approach the band edges. The electron trap due
to third neighboring B-H pairs is estimated to be 20~meV below the
conduction band bottom. Second neighbor pairs already show a clean
gap. Finally, the ground state of BH shows a true chemical passivation.
As proposed nearly three decades ago \citep{Pankove1985} -- it comprises
a covalently connected, fully saturated $\equiv\textrm{Si-H}\cdots\textrm{B}\equiv$
structure, where Si, B and H are four-fold, three-fold and mono coordinated,
respectively.

The above picture is essentially identical for pairs involving heavier
group-III species ($X$). From the electron localization function
analysis, we found that the ground state of all $X$H pairs also comprise
fully saturated $\equiv\textrm{Si-H}\cdots X\equiv$ structures. The
three-atom structure is linear for BH, whereas it is puckered for
other $X$H pairs with the Si-H bond making an angle of about 20°
away from the $\langle111\rangle$ direction.

Motion of H in the $\{111\}$ plane, via oscillations for the case
of BH, or rotation of the Si-H bond around $\langle111\rangle$ for
other $X$H pairs, is found to be governed by a very flat potential
energy landscape with variations of a few meV. This picture is compatible
with the model of Stavola \emph{et~al.} \citep{Stavola1987,Stavola1988a},
which describes the vibrational activity of the pairs as arising from
the coupling between a low-frequency degree of freedom (with tens
of cm$^{-1}$ involving atomic motion perpendicular to $\langle111\rangle$),
and a high-frequency stretching mode of the Si-H unit.

There are two main sources of error that affect the calculated frequencies:
neglected anharmonic regions of the potential energy surface, and
inherent insufficiencies in the electronic structure method for its
evaluation. High frequencies with well defined potential valleys can
be easily corrected by \emph{ad-hoc} shifts or scaling factors \citep{Merrick2007}.
A more elaborate method involves the extraction of anharmonic frequencies
directly from atomic trajectory data of first-principles molecular
dynamics at finite temperatures \citep{Wang2019}. On the other hand,
the low-frequency rotational motion of $X$H (and its coupling to
the Si-H stretching mode) belongs to a class of extreme cases, involving
a shallow potential with several minima, where the error bar of first-principles
electronic structure calculations overlaps many times the potential
variations and the separation between the rotovibrational levels.
Further progress in the understanding the low-frequency motion of
BH could however be attained by construction of the potential, partially
by first-principles and fitting to experimental data, and solving
the anharmonic Schrödinger equation. This approach has been successful
in the description of the rotovibrational motion of bond centered
interstitial oxygen in Si \citep{Kaneta2003,Kaneta1990,Lassmann2012}.

The calculated binding energies and dissociation barriers of $X$H
pairs are respectively in the range 0.76-0.95~eV and 1.18-1.37~eV.
They depend weakly on the acceptor species, with BH showing the smallest
figures and GaH showing an off-trend deviation with increasing the
mass of the acceptors. These features were previously observed \citep{Zundel1989}
and are now theoretically accounted for. We demonstrate that size/strain
effects are not essential to explain the observed trends. Instead,
they are justified based on the relatively stronger B-Si bond (that
has to be broken before formation of the BH pair), and the d-block
contraction of the gallium species.

Ground state $X$H pairs do not interact with minority carriers in
p-type Si. However, if $X^{-}$ and H$^{+}$ become separated by several
Si-Si bonds, the hydrogen atom could trap electrons, and that is found
to decrease the dissociation energy in two manners: (1) via reaction
$X^{-}\textrm{-H}^{+}+\textrm{e}^{-}\rightarrow X^{-}+\textrm{H}^{0}$,
which lowers $E_{\textrm{d}}$ by of up to 0.17~eV in comparison
to an electron deprived dissociation; (2) due to the accelerated migration
of H$^{0}$. The complex migration of H$^{0}$ makes the latter effect
more difficult to be quantified \citep{Gomes2022}. Both effects are
invoked to play a role in the enhanced injection-/photo-induced dissociation
of BH pairs by slowing the rate of the back reaction $X\textrm{B}\leftarrow X^{-}+\textrm{H}^{+}$
against that of the forward reaction during annealing treatments.

For the specific case of BH, the dissociation enhancement upon electron
trapping is likely to be limited by the relatively large barrier for
breaking the pair (before it can trap electrons), $E_{\textrm{d1}}\approx1$~eV.
This translates into a decrease of the dissociation energy by about
0.2~eV when compared to the same property without electron trapping.
For other acceptor-hydrogen pairs on the other hand, the barrier for
the first H jump is lower, the dissociation energy is now limited
by $E_{\textrm{d}}^{*}=E_{\textrm{b}}^{*}+E_{\textrm{m}}^{*}$ (see
Fig.~\ref{fig2}), and the effect of electron trapping is expected
to be even more conspicuous due to the fast escape of H$^{0}$. In
the extreme case of a negligible $E_{\textrm{m}}^{*}$, we would have
$E_{\textrm{d}}^{*}\geq0.8$~eV for the heavier $X$H pairs.

In Ref.~\citep{Fattah2022} it was postulated that regeneration of
BO-LID-degraded solar Si results from the transfer of H from BH pairs
to boron-dioxygen LID defects, effectively passivating the latter
and leading to a slightly more stable BO$_{2}$-H complex. Our results
indicate that the role of light in that process is to introduce a
population of electrons that promote a carrier-induced destabilization
of BH pairs in favor of more stable (and perhaps less light-sensitive)
BO$_{2}$-H complexes.

From the calculation of the free energy change across the reaction
$\textrm{B}^{-}+\textrm{H}^{+}\rightleftarrows\textrm{BH}$, we estimated
the annealing temperature of BH in the dark at about 180~°C. This
figure considered a boron concentration $[\textrm{B}]\sim10^{15}$~cm$^{-3}$.
The process is dominated by the larger configurational entropy of
the reactants, and due to its dependence on {[}B{]}, the annealing
temperature shows some variation (tens of degrees Celsius). We did
not find any significant impact of the hydrogen mass to the free energy
change of the reaction, and therefore to the annealing temperature.
Note that this conclusion refers to an equilibrium property, and it
is not contradictory with different dissociation rates of BH and BD
due to comparably more frequent H jumps.

The interaction of H$_{2}$ molecules with the acceptors, interstitial
oxygen and substitutional carbon was also investigated. In monocrystalline
p-type Si wafers, oxygen and carbon are abundant impurities and along
with the acceptors, they compete for trapping the molecules. In general,
for remote impurity-H$_{2}$ pairs (separation $r\gtrsim10$~Å) strain
interactions result in an energy change within $\pm10$~meV only.
Significant differences become noticeable only for $r\lesssim5$~Å,
where the magnitude of the potentials becomes larger than $k_{\textrm{B}}T$
at room temperature. Interstitial oxygen was found to be the most
attractive center for the formation of impurity-H$_{2}$ pairs (without
molecular dissociation). The calculated binding energy of adjacent
O-H$_{2}$ pairs exceeds 0.2~eV. Carbon is the most repulsive impurity.
Boron was also found to be repulsive for the molecules, whereas larger
acceptors are slightly attractive. These effects are well accounted
for by the relative volume that is available for H$_{2}$ in the vicinity
of the several impurities. Essentially, compressive (O, Ga, Al, In)
and tensile (C and B) defects, respectively increase and reduce the
volume of their first neighboring tetrahedral interstitial sites.
Hence, they respectively attract and repel the molecules before any
intimate reaction takes place.

The chemical reaction of H$_{2}$ molecules with boron, and subsequent
molecular dissociation, was found to involve the collision of H$_{2}$
with a first neighboring B-Si bond. An activation energy of 1.1~eV
was found for this process. This barrier is considerably lower than
the 1.62~eV which was previously estimated for H$_{2}$ dissociation
in pristine bulk Si \citep{Gomes2022}, but also larger than the barrier
for H$_{2}$ migration (0.78~eV \citep{Markevich1998,Pritchard1998}).
These findings offer an alternative explanation for H$_{2}$ dissociation
in B-doped Si without the involvement of holes, as proposed in Ref.~\citep{Voronkov2017}
for the first step of reaction \ref{eq:vfmodel} described in the
Introduction. We also found that immediately after the saddle point
for H$_{2}$ dissociation next to B, a metastable structure involving
H$^{-}$ next to BH is attained. From here, hole capture and reconfiguration
events are likely to influence the final result, which could be formation
of BH$_{2}^{-}$, BH$_{2}^{+}$ or BH pairs.

An analogous study was carried out for the dissociation of H$_{2}$
next to carbon impurities. The height of the barrier for $\textrm{C}+\textrm{H}_{2}\rightarrow\textrm{CH}_{2}$
was found to be $E_{\textrm{d}}=1.35$~eV. This suggests that C is
also a catalyst for H$_{2}$ dissociation in Si, although perhaps
less effective than boron. Along the reaction, an intermediate stable
structure made of close C-H$\cdots$H-Si bonds was found. However,
this state is electrically inactive, therefore the participation of
carriers is unlikely to occur during further transformations toward
the lowest energy CH$_{2}$ structures.

The most favorable products from the reaction of H$_{2}$ with B and
C are BH pairs and CH$_{2}$ complexes, respectively. These correspond
to respective potential energy drops of 1.19 eV and 0.91 eV per molecule.
After considering finite temperature effects, the free energy drop
across $\textrm{H}_{2}+2\textrm{B}^{-}+2\textrm{h}^{+}\rightarrow2\textrm{BH}$
was found to be still invariably larger than that for $\textrm{H}_{2}+\textrm{C}\rightarrow\textrm{CH}_{2}$.
From these results and the calculated reaction barriers, we conclude
that during the cooling of Si that has been in contact with a hot
hydrogen source, interstitial H$_{2}$ molecules are more likely to
react with boron than carbon, unless the concentration of the latter
is dominant.

The results point toward a strong thermodynamic drive for dispersion
of H in the form of BH pairs in B-doped Si. Dihydrogenated boron is
a by-product along the $2\textrm{B}^{-}+\textrm{H}_{2}+2\textrm{h}^{+}\rightarrow2\textrm{BH}$
reaction. BH$_{2}$ in Si is a negative-$U$ defect with a metastable
neutral state. The $(-/+)$ transition is anticipated at $E(-/+)-E_{\textrm{v}}=0.26$~eV
and involves different geometries for the donor and acceptor states.
The positive charge state has two Si-H bonds next to undercoordinated
boron, giving rise to an electron trap with an energy level at $E_{\textrm{c}}-0.19$~eV.
The negative charge state has a four-fold coordinated boron atom and
is responsible for a shallow acceptor level close to that of isolated
boron. As for heavier acceptor-H$_{2}$ complexes involving Al, Ga
and In, they mainly differ from BH$_{2}$ in that the donor states
are shallow effective-mass-like. As a corollary, besides the subtraction
of one hole, formation of heavier $X\textrm{H}_{2}^{+}$ complexes
lead to donation of one electron as well.

BH$_{2}^{+}$ and BH$_{2}^{-}$ are not very stable complexes. The
acceptor is not even stable above $T~\sim100$~K against $\textrm{B}^{-}+\textrm{H}_{2}$
in p-type Si. As for the donor state, it becomes less stable than
$\textrm{B}^{-}+2\textrm{H}^{+}$ at $T\gtrsim300$~K and the activation
energy for dissociation of a proton is about 0.8~eV. These results,
along with the calculated donor transition of BH$_{2}$ at $E_{\textrm{c}}-0.19$~eV,
indicate that BH$_{2}$ is effectively a donor defect (without a stable
acceptor) in the range $T\approx100\textrm{-}300$~K. They also support
the assignment of this defect to the experimentally detected electron
trap at $E_{\textrm{c}}-0.175$~eV, which was found to anneal out
at 280~K, and was connected to the LeTID of solar cells based on
B-doped Si \citep{Guzman2021}.

From the calculated local vibrational mode frequencies of boron-hydrogen
defects, including B- and H-related isotope shifts, we confirm the
structure of the BH pairs, and in particular the chemical passivation
model, according to which a Si-H covalent bond is established. We
have also calculated the frequency and relative intensity of several
vibrational modes of BH$_{2}$ complexes. These results are expected
to provide guidance for future experiments toward a better identification
of these centers.

Finally, regarding the LeTID of boron-doped Si solar cells, and in
the light of our results, we recall the sequence of reactions proposed
by de~Guzman \emph{et~al.} \citep{Guzman2021} for the degradation
and recovery steps under dark annealing conditions,

\[
\begin{array}{rcl}
\textrm{Latent} &  & \hspace{0.4cm}\textrm{H}_{2}+2\textrm{B}^{-}+2\textrm{h}^{+}\\
 &  & \hspace{1.4cm}\bigg\downarrow\hspace{0.2cm}(1)\:\:\tau,\rho\:\textrm{degradation}\\
\textrm{Degraded} &  & \hspace{0.5cm}\textrm{B}\textrm{H}_{2}^{+}(\textrm{D})+\textrm{B}^{-}\\
 &  & \hspace{1.4cm}\bigg\downarrow\hspace{0.2cm}(2)\:\:\tau\:\textrm{regeneration}\\
\textrm{Passivated} &  & \hspace{1.1cm}2\textrm{BH}\\
 &  & \hspace{1.4cm}\bigg\downarrow\hspace{0.2cm}(3)\:\:\rho\:\textrm{recovery}\\
\textrm{Recovered} &  & \textrm{H}_{2}(\textrm{sink})+2\textrm{B}^{-}+2\textrm{h}^{+}
\end{array}
\]

The above reaction sequence adds a \emph{Degraded} state to the three-state-model
of Voronkov and Falster \citep{Voronkov2017} (see Eq.~\ref{eq:vfmodel}),
which has been used to explain changes in resistivity of hydrogenated
boron doped Si crystals during heat/illumination treatments. We refer
to the three states of the original model as \emph{Latent}, \emph{Passivated}
and \emph{Recovered}. LeTID initiates from a Latent state, where H$_{2}$
molecules linger in the as-fired/quenched devices. Lifetime ($\tau$)
degradation occurs during step (1) when heat is provided, and H$_{2}$
is able to escape from trapping sites, most notably O impurities,
and react with B to form BH$_{2}$. We note that already at this stage,
for each BH$_{2}^{+}$ complex that is formed, two free-holes are
subtracted. This means that the resistivity ($\rho$) change and lifetime
degradation are concurrent, and any measured resistivity increase
rate, also reflects the lifetime degradation rate. The calculated
activation energy for step (1) is 1.1~eV and corresponds to the dissociation
of H$_{2}$ next to B. This quantity compares well with the LeTID
activation energy of 1.08 eV from the study of Vargas $et~al.$ \citep{Vargas2019}.

The recovery of lifetime and resistivity occur separately. Lifetime
regeneration involves the annealing of BH$_{2}^{+}(\textrm{D})$ and
formation of BH pairs in step (2) According to our calculations, the
activation energy for $\textrm{BH}_{2}^{+}(\textrm{D})\rightarrow\textrm{BH}+\textrm{H}^{+}$
is 0.8 eV and the proton should be captured by another boron atom.
Importantly, step (2) does not involve a change in the free-hole concentration,
but rather attaining a passivated state with a weak recombination
activity.

Following Refs.~\citep{Voronkov2017,Walter2022}, one may speculate
that the recovery of resistivity upon prolonged dark annealings, involves
the dissociation of BH pairs and subsequent trapping of hydrogen at
some kind of sink. This would explain the full recovery of the devices
and reactivation of the B dopants as described by step (3). The calculated
dissociation barrier of BH (1.2~eV) is close to the measured counterpart
(1.28~eV \citep{Zundel1989}) and also close to the activation energy
for the recovery of resistivity (1.1-1.3~eV \citep{Vargas2019,Winter2021,Acker2022}).

\noindent \appendix

\section{Configurational entropy for $\textrm{B}^{-}+\textrm{H}^{+}\rightarrow\textrm{BH}$\label{apdx:a}}

For the evaluation of configurational entropy, we have to estimate
the total number of equivalent microstates in the reactants and products
sides of the reaction of interest, $W_{\textrm{R}}$ and $W_{\textrm{P}}$
, respectively. This is only tractable upon consideration of a few
assumptions.

Firstly, we consider a crystalline sample made of $n_{\textrm{Si}}$
silicon sites, containing $n_{\textrm{B}}$ substitutional boron acceptors
and $n_{\textrm{H}}$ hydrogen atoms subject to $n_{\textrm{Si}}\gg n_{\textrm{B}}\gg n_{\textrm{H}}$;

Next, because hydrogen is a fast diffuser and boron is strongly \emph{anchored}
to the lattice, at low temperatures all hydrogen is inferred to be
trapped by boron (in the form of BH pairs), so that the number of
pairs is also $n_{\textrm{H}}$. The remaining boron is electrically
activated;

At high temperatures, all pairs dissociate into uncorrelated B$^{-}$
and H$^{+}$ elements. The number of free carriers is conserved along
the process and no change of electronic entropy is considered;

Finally, at low temperatures, BH pairing is only considered to occur
upon formation of first neighboring complexes. Any other configurations
are, for the sake of the calculation of configurational entropy, equivalent
to isolated B$^{-}$ and H$^{+}$ species.

The relevant reaction is therefore,

\begin{equation}
n_{\textrm{B}}\textrm{B}^{-}+n_{\textrm{H}}\textrm{H}{}^{+}\rightarrow(n_{\textrm{B}}-n_{\textrm{H}})\textrm{B}{}^{-}+n_{\textrm{H}}\textrm{BH}.
\end{equation}

When all BH are dissociated, the number of microstates for the reactants
side is found by first distributing the $n_{\textrm{B}}$ acceptors
(there are $n_{\textrm{Si}}!/(n_{\textrm{Si}}-n_{\textrm{B}})!n_{\textrm{B}}!$
ways to do that), and for each boron-related microstate we distribute
the $n_{\textrm{H}}$ hydrogen atoms among the available $2n_{\textrm{Si}}-4n_{\textrm{B}}$
bond center sites not adjacent to boron. The total number of ways
to perform the two operations is,

\begin{equation}
W_{\textrm{R}}=\frac{n_{\textrm{Si}}!}{(n_{\textrm{Si}}-n_{\textrm{B}})!n_{\textrm{B}}!}\times\frac{(2n_{\textrm{Si}}-4n_{\textrm{B}})!}{(2n_{\textrm{Si}}-4n_{\textrm{B}}-n_{\textrm{H}})!n_{\textrm{H}}!}.
\end{equation}

On the other hand, when all H is trapped by boron, the total number
of possible microstates for the products side is given by the number
of ways to distribute the $n_{\textrm{B}}$ boron atoms times the
number of ways to distribute $n_{\textrm{H}}$ hydrogen atoms on $4n_{\textrm{B}}$
bond centered sites next to boron,

\begin{equation}
W_{\textrm{P}}=\frac{n_{\textrm{Si}}!}{(n_{\textrm{Si}}-n_{\textrm{B}})!n_{\textrm{B}}!}\times\frac{(4n_{\textrm{B}})!}{(4n_{\textrm{B}}-n_{\textrm{H}})!n_{\textrm{H}}!}.\label{eq:wp-apdxa}
\end{equation}

Using Stirling’s approximation, we find that the total entropy change
is linear with $n_{\textrm{H}}$ , and we may derive the change of
configurational entropy per BH pair as,

\begin{equation}
\Delta S_{\textrm{conf}}=\frac{k_{\textrm{B}}}{n_{\textrm{H}}}\ln\left(\frac{W_{\textrm{P}}}{W_{\textrm{R}}}\right)=k_{\textrm{B}}\ln(2c_{\textrm{B}}),
\end{equation}
where $k_{\textrm{B}}$ is the Boltzmann constant and $c_{\textrm{B}}=n_{\textrm{B}}/n_{\textrm{Si}}$
is the fractional concentration of boron. If we were to consider up
to second neighboring bonds to distinguish between pairing formation/dissociation,
the factor of 2 next to $c_{\textrm{B}}$ would have to be replaced
by 8, thus having a minute impact on the final result.

\section{Configurational entropy for $\textrm{B}^{-}+\tfrac{1}{2}\textrm{H}_{2}+\textrm{h}^{+}\rightarrow\textrm{BH}$\label{apdx:b}}

Now we implicitly assume that multi-trapping of H at B is negligible,
so that the number of resulting pairs is also $n_{\textrm{H}}$, and
that after the reaction, the remaining $(n_{\textrm{B}}-n_{\textrm{H}})$
boron ions are ionized. Again, taking $n_{\textrm{Si}}\gg n_{\textrm{B}}\gg n_{\textrm{H}}$,
the relevant reaction is,

\begin{equation}
n_{\textrm{B}}(\textrm{B}^{-}+\textrm{h}^{+})+\frac{n_{\textrm{H}}}{2}\textrm{H}_{2}\rightarrow(n_{\textrm{B}}-n_{\textrm{H}})(\textrm{B}^{-}+\textrm{h}^{+})+n_{\textrm{H}}\textrm{BH},\label{eq:hdis-anxb}
\end{equation}
which highlights the fact that each hydrogen molecule is capable of
subtracting two holes from the sample. Since the number of free holes
is not conserved across reaction \ref{eq:hdis-anxb}, the effect of
electronic entropy should be assessed as well. In this appendix we
evaluate the configurational term.

For the reactants side, we assume that the $n_{\textrm{B}}$ boron
ions and the $n_{\textrm{H}}/2$ molecules are perfectly uncorrelated,
and they can occupy any of the $n_{\textrm{Si}}$ substitutional and
$n_{\textrm{Si}}$ tetrahedral interstitial sites, respectively. The
number of ways to distribute such populations is

\begin{equation}
W_{\textrm{R}}=\frac{n_{\textrm{Si}}!}{(n_{\textrm{Si}}-n_{\textrm{B}})!n_{\textrm{B}}!}\times\frac{2^{n_{\textrm{H}}/2}n_{\textrm{Si}}!}{(n_{\textrm{Si}}-n_{\textrm{H}}/2)!(n_{\textrm{H}}/2)!},\label{eq:apdx-b-h2}
\end{equation}
where the factor of $2^{n_{\textrm{H}}/2}$ accounts for the two possibilities
that we have to form each molecule.

As for the number of microstates in the products side ($W_{\textrm{P}}$)
we can use Eq.~\ref{eq:wp-apdxa} from Appendix~\ref{apdx:a}. Hence,
the configurational entropy change per BH pair can be approximated
to

\begin{equation}
\Delta S_{\textrm{conf}}=\frac{k_{\textrm{B}}}{n_{\textrm{H}}}\ln\left(\frac{W_{\textrm{P}}}{W_{\textrm{R}}}\right)=\frac{k_{\textrm{B}}}{2}\left[\ln\left(\frac{8c_{\textrm{B}}^{2}}{c_{\textrm{H}}}\right)+1\right],
\end{equation}
where $c_{\textrm{B}}=n_{\textrm{B}}/n_{\textrm{Si}}$ and $c_{\textrm{H}}=n_{\textrm{H}}/n_{\textrm{Si}}$
are the fractional concentrations of boron and hydrogen, respectively.

\section{Configurational entropy for $\textrm{C}+\textrm{H}_{2}\rightarrow\textrm{CH}_{2}$\label{apdx:c}}

It is known that CH pairs are not very stable. They dissociate at
$T\sim100$~°C in darkness and even below 230~K upon illumination
with above band gap light \citep{Yoneta1991}. Therefore, we assume
that interactions between H$_{2}$ and C upon quenching from H$_{2}$
exposure at high temperatures, essentially lead to formation of CH$_{2}$
complexes \citep{Markevich2001}.

Again, taking $n_{\textrm{Si}}\gg n_{\textrm{C}}\gg n_{\textrm{H}}$,
where $n_{\textrm{C}}$ is the number of substitutional carbon impurities
in the sample, and the number of resulting CH$_{2}$ complexes as
$n_{\textrm{H}}/2$, the relevant reaction is,

\begin{equation}
n_{\textrm{C}}\textrm{C}+(n_{\textrm{H}}/2)\textrm{H}_{2}\rightarrow(n_{\textrm{C}}-n_{\textrm{H}}/2)\textrm{C}+(n_{\textrm{H}}/2)\textrm{CH}_{2},
\end{equation}

On the reactants side, we have $n_{\textrm{Si}}!/(n_{\textrm{Si}}-n_{\textrm{C}})!n_{\textrm{C}}!$
ways to distribute the C atoms while H$_{2}$ molecules can occupy
tetrahedral interstitial sites. Considering also that there are two
ways to form each molecule (see Eq.~\ref{eq:apdx-b-h2}), the total
number of microstates becomes,

\begin{equation}
W_{\textrm{R}}=\frac{n_{\textrm{Si}}!}{(n_{\textrm{Si}}-n_{\textrm{C}})!n_{\textrm{C}}!}\times\frac{2^{n_{\textrm{H}}/2}n_{\textrm{Si}}!}{(n_{\textrm{Si}}-n_{\textrm{H}}/2)!(n_{\textrm{H}}/2)!}.
\end{equation}

Conversely, when all H$_{2}$ is trapped at the C atoms, the total
number of possible microstates for the products side is now the number
of ways to distribute the $n_{\textrm{C}}$ carbon atoms times the
number of ways to combine them with the $n_{\textrm{H}}/2$ pairs
of H atoms. Considering that each CH$_{2}$ complex has four-fold
orientational degeneracy, and again the two possibilities for the
formation of each complex (due to the existence of two H atoms), one
finds,

\begin{equation}
W_{\textrm{P}}=\frac{n_{\textrm{Si}}!}{(n_{\textrm{Si}}-n_{\textrm{C}})!n_{\textrm{C}}!}\times\frac{2^{n_{\textrm{H}}/2}(4n_{\textrm{C}})!}{(4n_{\textrm{B}}-n_{\textrm{H}}/2)!(n_{\textrm{H}}/2)!}.
\end{equation}

Following the same procedure as in Appendix~\ref{apdx:a} and \ref{apdx:b},
we have the configurational entropy change per H atom,

\begin{equation}
\Delta S_{\textrm{conf}}=\frac{k_{\textrm{B}}}{2}\ln\left(4c_{\textrm{C}}\right),
\end{equation}
where $c_{\textrm{C}}=n_{\textrm{C}}/n_{\textrm{Si}}$ is the fractional
concentrations carbon.

\section{Electronic free energy\label{apdx:d}}

For solid state reactions where there is a variation in the number
of free carriers, the contribution of electronic entropy to the free
energy change can be comparable to other terms. This is particularly
relevant when we are dealing with reactions that involve the thermal
promotion of carriers, such as $\textrm{D}^{0}\rightarrow\textrm{A}^{-}+\textrm{h}^{+}$,
where D is a deep or electronically inactive defect, while A stands
for an arbitrary shallow acceptor, and $\textrm{h}^{+}$ a free hole
in thermal equilibrium with the sample.

We restrict our analysis to the low doping regime at room temperature
and up to few hundred Kelvin (LeTID conditions), where virtually all
shallow acceptors (boron or BH$_{2}$(A) complexes) are ionized. At
these temperatures, intrinsic excitations are negligible and that
is reflected in the excellent agreement between the calculated specific
heat, obtained from vibrational degrees of freedom only, and the measurements
\citep{Estreicher2004,Gomes2022}. For low doping conditions, the
effective density of valence band states largely exceeds the free
hole concentration ($N_{\textrm{v}}\gg p$) and we can apply Boltzmann
statistics. For instance, for the reaction

\begin{equation}
\textrm{B}^{-}+\tfrac{1}{2}\textrm{H}_{2}+\textrm{h}^{+}\rightarrow\textrm{BH},
\end{equation}
the electronic free energy change is estimated from that of a free-hole
gas with density $\Delta p=fc_{\textrm{H}}[\textrm{Si}]$, where $f=\exp(-E_{\textrm{h}}/k_{\textrm{B}}T)$
is the Boltzmann distribution function, and $E_{\textrm{h}}$ the
hole binding energy to the acceptor ($E_{\textrm{h}}=46$~meV for
the boron acceptor \citep{Ramdas1981}).

In the classical limit, and as proposed by Estreicher \emph{et~al.}
\citep{Estreicher2004}, the electronic free energy change due to
thermal release of a hole from a shallow state into the electronic
thermal bath of the sample, is given by \citep{Kubo1988}

\begin{equation}
\Delta F_{\textrm{elec,h}}=f(\mu-k_{\textrm{B}}T),\label{eq:felec-apdxd}
\end{equation}
where $\mu$ is the electronic chemical potential with respect to
the valence band top. Equation \ref{eq:felec-apdxd} finds the Helmholtz
free energy by simply subtracting $pV$ (which is $k_{\textrm{B}}T$
for a classical gas of non-interacting holes) to the Gibbs free per
hole ($\mu$).

The chemical potential is approximated to $\mu=k_{\textrm{B}}T\ln(\Delta p/N_{\textrm{v}})$,
where $N_{\textrm{v}}$ is obtained numerically in the range $T=200\textrm{-}500$~K
as $N_{\textrm{v}}=3.10\times10^{19}(T/300)^{1.85}$~cm$^{-3}$ \citep{Green1990}.
For the temperatures of interest, i.e., $T\approx300\textrm{-}500$~K,
we have $f\approx1$ and provided that $\Delta p\ll p\ll N_{\textrm{v}}$,
equation~\ref{eq:felec-apdxd} allows us to estimate the electronic
free energy variation due to small increase of free holes, across
a reaction that involves the full ionization/passivation of shallow
acceptors.
\begin{acknowledgments}
We acknowledge the FCT through projects LA/P/0037/2020, UIDB/50025/2020,
UIDP/50025/2020 and 2021.09643.CPCA (Advanced Computing Project using
the Oblivion supercomputer). The work in the UK was funded by EPSRC
via grant EP/TO25131/1.
\end{acknowledgments}

\bibliographystyle{apsrev4-2}
%\bibliography{refs}


\end{document}